\documentclass[useAMS,usenatbib]{mn2e}
\usepackage{graphicx}
\usepackage{amsmath}
\usepackage{color}

\def\be{\begin{equation}}
\def\ee{\end{equation}}
\def\ba{\begin{eqnarray}}
\def\ea{\end{eqnarray}}
\def\go{\mathrel{\raise.3ex\hbox{$>$}\mkern-14mu
             \lower0.6ex\hbox{$\sim$}}}
\def\lo{\mathrel{\raise.3ex\hbox{$<$}\mkern-14mu
             \lower0.6ex\hbox{$\sim$}}}

\newcommand{\hatl}{\hat{\mbox{\boldmath $l$}}}

\def\bomega{{\mbox{\boldmath $\omega$}}}

\def\cN{{\cal N}}
\def\bN{{\bf N}}

\def\bcN{{\mbox{\boldmath ${\cal N}$}}}
\def\rin{r_{\rm in}}

\begin{document}
\title[Spin Evolution of Accreting Magnetic Protostars]
{Evolution of Spin Direction of Accreting Magnetic Protostars
and Spin-Orbit Misalignment in Exoplanetary Systems: II. Warped Discs}
\author[F. Foucart and D. Lai]
{Francois Foucart$^{1}$\thanks{Email: fvf2@cornell.edu}
and Dong Lai$^{1,2}$\footnotemark[1] \\
$^1$Center for Space Research,
Department of Astronomy, Cornell University, Ithaca, NY 14853, USA \\
$^2$Kavli Institute for Theoretical Physics,
University of California, Santa Barbara, CA 93106, USA\\}

\pagerange{\pageref{firstpage}--\pageref{lastpage}} \pubyear{2010}

\label{firstpage}
\maketitle

\begin{abstract}
Magnetic interactions between a protostar and its accretion disc can
induce warping in the disc and produce secular changes in the
stellar spin direction, so that the spin axis may not always be perpendicular to the disc.
This may help explain the 7-degree misalignment between the ecliptic plane of the solar system
and the sun's equatorial planem as well as play a role in producing the recently observed 
spin-orbit misalignment in a number of exoplanetary systems.
We study the dynamics of warped protoplanetary
discs under the combined effects of magnetic warping/precession
torques and internal stresses in the disc, including viscous damping
of warps and propagation of bending waves.  We show that when the
outer disc axis is misaligned with the stellar spin axis, the disc
evolves towards a warped steady-state on a timescale that depends on
the disc viscosity or the bending wave propagation speed, but in all
cases is much shorter than the timescale for the spin evolution (of order of a 
million years). Moreover, for the most likely physical parameters characterizing 
magnetic protostars, circumstellar discs
and their interactions, the steady-state disc, averaged over the stellar rotation period, has a 
rather small warp such that
the whole disc lies approximately in a single plane determined by the outer disc boundary
conditions, although 
more extreme parameters may give rise to larger disc warps. In agreement with our
recent analysis \citep{lfl1} based on flat discs, we
find that the back-reaction magnetic torques of the slightly warped disc on the
star can either align the stellar spin axis with the disc axis or
push it towards misalignment, depending on the parameters of the
star-disc
system. This implies that newly formed planetary systems may have
a range of inclination angles between the stellar spin axis and the
orbital angular momentum axis of the planetary orbits. 
\end{abstract}


\section{Introduction}

In a recent paper [\citet{lfl1}, hereafter Paper I], we
proposed a novel mechanism for producing misalignment between the spin
axis of a protostar and the normal vector of its circumstellar
disc. Our work was motivated by recent measurements of the
sky-projected stellar obliquity using the Rossiter-McLaughlin effect
in transiting exoplanetary systems, which showed that a large fraction
of the systems containing hot Jupiters have misaligned stellar spin
with respect to the planetary angular momentum axis [see \citet{triaud};
\citet{win3} and references therein]. Additional evidence for
nonzero stellar obliquity came from the statistical analysis of the
apparent rotational velocities ($v\sin i_\star$) of planet-bearing
stars \citep{sch1}.

The basic mechanism (``Magnetically driven misalignment'') for
producing spin -- disc misalignment in accreting protostellar systems
can be sumarized as follows (Paper I).  The magnetic field of a
protostar (with $B_\star\go 10^3$~G) penetrates the inner region of
its accretion disc. These field lines link the star and the disc in a
quasi-cyclic fashion (e.g., magnetic field inflation followed by
reconnection; see~\cite{bou} and~\cite{ale} for observational evidence).  Differential rotation between the star and the disc
not only leads to the usual magnetic braking torque on the disc, but
also a warping torque which tends to push the normal axis of the
inner disc away from the spin axis\footnote{The warping torque vanishes
if the angular momentum of the disc is exactly aligned with the stellar spin, but exists for 
arbitrary small angles. A flat disc in the aligned configuration is in an unstable equilibrium}. Hydrodynamical
stresses in the disc, on the other hand, tend to inhibit significant disc warping.  The
result is that, for a given disc orientation imposed at large radii
(e.g., by the angular momentum of the accreting gas falling onto the
disc), the back-reaction of the warping torque can push the stellar
spin axis toward misalignment with respect to the disc normal vector.
Planets formed in the disc will then have a misaligned
orbital normal axis relative to the stellar spin axis, assuming that
no evolution mechanism occurring after the dissipation of the disc forces the alignment of the system.

The process of planetary system formation can be roughly divided into
two stages \citep{jt1}. In the first stage, which lasts a
few million years until the dissipation of the gaseous protoplanetary
disc, planets are formed and undergo migration due to tidal
interactions with the gaseous disc [\citet{lin1}; see \citet{pap1} for a
review]. The second stage, which lasts from when the disc has
dissipated to the present, involves dynamical gravitational
interactions between multiple planets, if they are produced in the
first stage in a sufficiently close-packed configuration \citep{jt1,cha1},
and/or secular interactions with a distant
planet or stellar companion \citep{eke1,wm1,ft1,wu1,nib1}.
The eccentricity distribution of exoplanetary systems and
the recent observational results on the spin -- orbit misalignment
suggest that the physical processes in the second stage
play an important role in determining the properties of
exoplanetary systems. Nevertheless, the importance of the first
stage cannot be neglected as it sets the initial condition for the possible
evolution in the second stage.
Our result in Paper I shows that at the end of the first stage,
the symmetry axis of the planetary orbit may be inclined with respect to
the stellar spin axis.

At first sight, it may seem strange that the magnetic field effects
can drive the stellar spin axis toward misalignment with respect to
the disc symmetry axis, given that the spin angular momentum of the
star ultimately comes from the disc and the disc contains a large
reservoir of angular momentum. The key to understand this is to
realize that when the gas reaches the magnetosphere boundary, its
angular momentum is much smaller than in the outer disc
(the specific angular momentum of the disc is $j_{\rm disc}(r) = \sqrt{GMr}$ 
for a Keplerian disc), and any magnetic torque, which in
general can break the axisymmetry of the system, is of the same order
of magnitude as the accretion torque on the star.

A key assumption adopted in Paper I for the calculation of
the magnetic torque on the star from the disc is that the disc is flat.
This is a nontrivial assumption. Indeed, the magnetic coupling between the
star and the disc operates only in the innermost disc region (e.g., between
the inner radius $\rin$ and $r_{\rm int} \approx 1.5\rin$), and this region has a much
smaller moment of inertia than the star. Therefore, if there were no coupling
between this inner disc region and the outer disc,
the inner disc would be significantly warped on a timescale much shorter
than the timescale for changing the stellar spin
\citep{pl1}. If there is any secular
change in the stellar spin direction, the inner disc warp would then follow
the varying spin axis.
Clearly, in order to determine the long-term spin evolution of the star,
it is important to understand the dynamics of the warped disc, taking into
account the magnetic torques on the inner disc and the hydrodynamical
coupling between different disc regions. This is the goal of our paper.


To be more specific, there is a hierarchy of timescales related to the combined
evolution of the stellar spin and the disc warp: 

(i) The dynamical time $t_{\rm dyn}$
associated with the spin frequency $\omega_s$, disc rotation frequency $\Omega$ and
the beat frequency $|\omega_s-\Omega|$. This is much shorter than the effects 
(steady-state disc warping and spin evolution) we study in this paper.

(ii) The warping/precession timescale of the inner disc 
[see Eq.~(\ref{eqn:Gamma_w})]
\ba
&&t_w\sim \Gamma_w^{-1}=\left(92\,{\rm days}\right) \left({1~{\rm kG}\over
  B_\star}\right)^{\!2} \left({2R_\odot\over R_\star}\right)^{\!6}
\left({M_\star\over 1\,M_\odot}\right)^{\!1/2}\nonumber\\
&&\qquad
\times \left({\rin\over 8R_\odot}\right)^{\!11/2}
\left({\Sigma\over 10\,{\rm g\,cm}^{-2}}\right)
\left(\zeta\cos\theta_\star\right)^{-1},
\label{eq:twarp}\ea
where $M_\star,\,R_\star,\,B_\star$ are the mass, radius and surface (dipole)
magnetic field of the protostar, respectively, $\theta_\star$ is the 
inclination angle of the stellar dipole relative to the spin, 
$\Sigma$ is the disc surface density, and $\zeta$ is a dimensionless magnetic twist parameter
of order unity related to the strength of the azimuthal magnetic field generated by star-disc twist.

(iii) The disc warp evolution timescale $t_{\rm disc}$. This is the time for the 
disc to reach a steady-state under the combined effects of magnetic torques
and internal fluid stresses (see Section 5). For high-viscosity discs, $t_{\rm disc}$ 
is the viscous diffusion time for the disc warp [see Eq.~(\ref{eq:tvis})] and depends
on the viscosity parameter $\alpha$ and the disc thickness $\delta=H/r$:
\be
t_{\rm vis}\sim (3000\,{\rm yrs})\left({\alpha\over 0.1}\right)
\left({\delta\over 0.1}\right)^{-2}\!\!\left({r\over 100\,{\rm AU}}\right)^{3/2}.
\ee
For low-viscosity discs ($\alpha\lo \delta$), $t_{\rm disc}$ is the propagation time
of bending waves across the whole disc and depends on the sound speed.
In general, $t_{\rm disc}$ can be several orders of magnitude larger than 
$t_w$.

(iv) The stellar spin evolution timescale. The magnetic misalignment 
torque on the star is of order $\mu^2/r_{\rm in}^3$ ($\mu$ being the magnetic dipole moment of
the star), which is comparable to the fiducial accretion torque, given for Keplerian discs by $\cN_0=\dot M\sqrt{GM_\star\rin}$. Assuming the spin angular momentum 
$J_s=0.2M_\star R_\star^2\omega_s$ (the value for a $\Gamma=5/3$ polytrope,
representing a convective star), we find the spin evolution time 
\ba
&&t_{\rm spin}={J_s\over \cN_0}
=(1.25\,{\rm Myr})\left(\!{M_\star\over 1\,M_\odot}\!\right)
\!\!\left({{\dot M}\over 10^{-8}{M_\odot}{\rm yr}^{-1}}\right)^{\!-1}
\nonumber\\
&&\qquad \times \left(\!{\rin\over 4R_\star}\!\right)^{\!-2}
\!\!{\omega_s\over\Omega(\rin)}.
\label{eq:tspin}\ea
In general $t_{\rm spin}\gg t_{\rm disc}$. In this paper we will study 
the evolution of the disc warp on timescales ranging from $t_w$ to $t_{\rm disc}$,
and the evolution of the stellar spin direction on timescales of order $t_{\rm spin}$.

It is important to note that we are not interested
in disc warpings that vary on the dynamical timescale $t_{\rm dyn}$ in this paper.
In general, when the stellar dipole axis is inclined with respect to the
spin axis, there will be periodic vertical forces at the rotation
frequency of the star acting on the inner disc\footnote{The forcing frequency
may also be twice of the spin frequency under certain conditions
(e.g., when the disc is partially diamagnetic); see \citet{lz1}.}.
These periodic forces will lead to the warping of the disc, particularly
for low-viscosity discs in which bending waves propagate
\citep{tp1,lz1}. Indeed, there
is observational evidence for such magnetically-warped discs.
For example, the recurrent luminosity dips observed in the classical T Tauri
star AA Tauri has been attributed to the periodic occultation of the central
star by a warped inner disc \citep{bou}.
However, such dynamical disc warps average exactly to zero over a rotation
period and have no effect on the secular evolution of the system.


The remainder of the paper is organized as follows. In Section
\ref{sec:analytic}, we summarize our analytical model of magnetopshere
-- disc interaction and derive the equation for the evolution of the
stellar spin axis when the disc is warped.  In Section 3 we present
theoretical formalisms for determining the steady-state and time
evolution of warped discs, for both high-viscosity regime (where warps
propagate diffusively) and low-viscosity regime (where warps propagate
as bending waves). An approximate analytical expression for the
steady-state linear warp is also derived (see Section 3.2.2).
In Section 4 we present numerical results for the steady-state disc warp profiles 
under various conditions and in Section 5 we study the time evolution of disc warps.
We examine in Section 6 how the inner disc warp and the stellar spin evolution
respond to variations of the outer disc, and discuss in Section 7 how this could, in principle, lead
to anti-aligned planetary orbits, even for discs with initial angular momentum nearly aligned
with the stellar spin. We conclude in Section 8 with 
a discussion of our results.

\section{Analytic Model of the Disc -- Magnetic Star System}
\label{sec:analytic}

\subsection{Magnetic Torques on the Disc}

The interaction between a magnetic star and a disc is complex
(see references in Paper I).
However, the key physical effects of this interaction on the disc
can be described robustly in a parametrized manner. The model
used throughout this paper is detailed in Paper I. Here, we will limit
ourselves to a brief summary of the magnetic torques acting on the disc.
 
The stellar magnetic field disrupts the accretion disc at the 
magnetospheric boundary, where the magnetic and plasma stresses balance.
For a dipolar magnetic field with magnetic moment $\mu$, we have
\be
\rin=\eta \left({\mu^4\over GM_\star\dot M^2}\right)^{1/7},
\label{alfven}\ee
where $\eta$ is a dimensionless constant somewhat less than unity ($\eta
\sim 0.5$ according to recent numerical simulations; see Long et
al.~2005~\footnote{In the notation of Long et al., $\eta=k_A/2^{1/7}$ and $k_A=1$
corresponds to the solution for spherical accretion}).  We take $\rin$ to be the inner edge
 of the disc. Before
being disrupted, the disc generally experiences nontrivial magnetic
torques from the star (Lai 1999; Paper I).  
Consider a cylindrical coordinate
system $(r,\phi,z)$, with the vertical axis Oz orthogonal to the plane of the disc.
The magnetic torques are of two types:
(i) A warping torque ${\bf N}_w$ which acts in a small interaction
region $\rin < r < r_{\rm int}$, where some of the stellar field
lines are linked to the disc in a quasi-cyclic fashion (involving
field inflation and reconnection).
These field lines are twisted by
the differential rotation between the star and the disc, generating
a toroidal field $\Delta B_\phi=\mp \zeta B_z^{(s)}$ from the quasi-static
vertical field $B_z^{(s)}$ threading the disc, where $\zeta\sim 1$~\citep{aly2,lov} and
the upper/lower sign refers to the value above/below the disc
plane. Since the toroidal field from the stellar dipole
$B_\phi^{(\mu)}$ is the same on both sides of the disc plane, the net
toroidal field $B_\phi=B_\phi^{(\mu)}+\Delta B_\phi$ differs above and below the disc plane, giving rise to
a vertical force on the disc. While the mean force (averaging over the
azimuthal direction) is zero, the uneven distribution of the force
induces a net warping torque which tends to push the orientation of
the disc angular momentum $\hatl$ away from the stellar spin axis
$\hat{\bomega}_s$ (see Paper I for a simple model for this effect, involving a metal plane in an
external magnetic field).  (ii) A precessional torque ${\bf N}_p$ which
arises from the screening of the azimuthal  electric current induced in the
highly conducting disc. This results in a difference in the radial
component of the net magnetic field above and below the disc plane
and therefore in a vertical force on the disc.  The resulting
torque tends to cause $\hatl$ to precess around
$\hat\bomega_s$. In Paper I, we parametrized the two magnetic torques 
(per unit area) on the disc as
\ba
{\bf N}_w &=& -(\Sigma r^2\Omega)\cos\beta\,
\Gamma_w \,\hatl\times(\hat{\bomega}_s\times\hatl),\label{eq:torquew}\\
{\bf N}_p&=&(\Sigma r^2\Omega)\cos\beta\,
\Omega_p \,\hat{\bomega}_s\times\hatl, 
\label{eq:torque}\ea
where $\Sigma(r)$ is the surface density, $\Omega(r)$ the 
rotation rate of the disc, and $\beta(r)$ is the disc tilt angle (the angle
between $\hatl (r)$ and the spin axis $\hat\bomega_s$).
The warping rate and precession angular frequency at radius $r$ are 
given by
\ba
&&\Gamma_w (r)=\frac{\zeta\mu^2}{4\pi r^7\Omega(r)\Sigma(r)}\cos^2\theta_\star,
\label{eqn:Gamma_w}\\
&&\Omega_p (r)=\frac{\mu^2}{\pi^2 r^7\Omega(r)\Sigma(r)
  D(r)}F(\theta_\star),
\label{eqn:Omega_p}
\ea
where $\theta_\star$ is the angle between the magnetic dipole axis and the 
spin axis, and the dimensionless function $D(r)$ is given by
\begin{equation}
D(r)={\rm max}~\left(\sqrt{r^2/r^2_{\rm in}-1}, \sqrt{2H(r)/r_{\rm in}}\right).
\label{eqn:D(r)}
\ee
with $H(r)$ the half-thickness of the disc.
The function $F(\theta_\star)$ depends on the dielectric properties of the disc.
We can write
\begin{equation}
F(\theta_\star)=2f\cos^2\theta_\star-\sin^2\theta_\star.
\end{equation}
If the stellar vertical field is entirely 
screened out by the disc, the parameter $f=1$; if only the time-varying component
of that field is screened out, we get $f=0$. In reality, $f$ lies 
between 0 and 1. 

The magnetic torque formulae given above
contain uncertain parameters (e.g., $\zeta$, which
parametrizes the amount of azimuthal twist of the magnetic field
threading the disc); this is inevitable given the complicated nature
of magnetic field -- disc interactions.
Also, while the expression for the
warping torque [eq.~(\ref{eq:torquew})]
is formally valid for large disc warps, the expression for
the precessional torque was derived under the assumption that the disc is
locally flat [eq.~(\ref{eqn:Omega_p}) is strictly valid only for a
completely flat disc \citep{aly1}]; when this assumption breaks
down (i.e., when $|\partial\hatl/\partial\ln r|$ is large), we expect
a similar torque expression to hold, but with modified numerical
factors (e.g. the function $D(r)$ in eq.~(\ref{eqn:Omega_p}) will be
different).  In the application discussed in the following sections,
we find that the condition $|\partial\hatl/\partial\ln r| \lo 1$ is always
satisfied.  Thus we believe that our simple formulae
capture the qualitative behavior of accretion discs subject to 
magnetic torques.

It is also worth noting that the expressions~(\ref{eq:torquew}-\ref{eq:torque}) for the
torques only correspond to the zero-frequency component of the magnetic forces acting on the
disc. The time varying components of these forces can also have significant effects. In particular, 
\citet{lz1} discussed how the components of the magnetic forces varying at the stellar spin frequency and at twice that frequency can excite bending waves in discs, while \citet{tp1} showed
that if the star has a dipole field misaligned with its rotation axis, magnetic effects create a steady-state warp in a frame corotating with the star. However, these ``dynamical waves'' 
average to zero over the stellar rotation period and do not affect the secular evolution of the stellar spin. In this paper, we concern ourselves only with long-term effects, effectively studying a disc profile averaged over multiple stellar rotations.

\subsection{Spin Evolution of the Star}
\label{sec:spinevol}

The effects of the magnetic torques on the
evolution of the star -- disc system are twofold. First, they will
cause the orientation of the disc $\hatl(r)$ to deviate from a flat
disc profile $\hatl(r)=\hatl_{\rm out}=\hatl(r_{\rm out})$, set at the
outer disc radius $r_{\rm out}$.  These deviations will be studied in
details for different disc parameters in Sections 3-5.
Second, the back-reaction of the torques will
change the orientation of the stellar spin axis on a longer timescale.
The secular evolution of the stellar spin under the combined
effects of matter accretion and star -- disc interactions is
explored in Paper I in the case of flat discs. Here we 
generalize the basic formulae derived in Paper I to warped discs.

In general, the spin angular momentum of the star, $J_s\hat\bomega_s$, evolves
according to the equation
\be
{d\over dt}\left(J_s\hat\bomega_s\right)=\bcN=
\bcN_l+\bcN_s+\bcN_w+\bcN_p.
\label{spin}\ee
Here $\bcN_l$ represents the torque component that is aligned with the
inner disc axis $\hatl(r_{\rm in})=\hatl_{\rm in}$. We 
parametrize $\bcN_l$ by
\be
\bcN_l=\lambda\dot M (GM_\star\rin)^{1/2}\,\hatl_{\rm in}
=\lambda \cN_0\,\hatl_{\rm in},
\label{eq:Nl}\ee
Equation~(\ref{eq:Nl}) includes
not only the accretion torque carried by the accreting
gas onto the star, $\dot M_{\rm acc} (GMr_{\rm in})^{1/2}\hatl$ (where
$\dot M_{\rm acc}$ may be smaller than $\dot M$, the disc accretion rate), but
also the magnetic braking torque associated with
the disc -- star linkage, as well as any angular momentum
carried away by the wind from the magnetosphere boundary~\citep{shu,rom1}.
All these effects are parametrized by the parameter $\lambda \lo 1$. In particular,
if a wind carries away most of the angular momentum of the inner disc, we may get
$\lambda \ll 1$.

The term $\bcN_s=-|{\cal N}_s| \hat\bomega_s$ represents a spindown torque
carried by a wind/jet from the open field lines region of the star (e.g. Matt \& Pudritz 2005).
The terms $\bcN_w$ and $\bcN_p$ represent the back-reactions
of the warping and precessional torques:
\be
\bcN_{w,p}=-\int_{\rin}^{r_{\rm out}}\! 2\pi r {\bf N}_{w,p}\,dr.
\ee
Since both $\bN_w$ and $\bN_p$ decrease rapidly with radius (as $r^{-5}$),
the integral can be carried out approximately, giving
\be
\bcN_p+\bcN_w \approx \cN_0\left[n_p\hat\bomega_s\times\hatl_{\rm in}
+n_w\hatl_{\rm in}\times (\hat\bomega_s\times\hatl_{\rm in})\right],
\label{eq:Nev}\ee
with 
\ba
&&n_p=-{4\over 3}{1 \over \pi\eta^{7/2}}F(\theta_\star)\,\cos\beta_{\rm in}\
,\label{eq:np}\\
&&n_w={\zeta [1-(\rin/r_{\rm int})^3]
\over 6\eta^{7/2}}\cos^2\!\theta_\star\,\cos\beta_{\rm in},
\ea
where $\cos\beta_{\rm in}=\hat\bomega_s\cdot\hatl_{\rm in}$. Note that both $\bcN_0 n_p$ 
and $\bcN_0 n_w$ are of order $\mu^2/r_{\rm in}^3$.

For a fixed outer disc orientation $\hatl_{\rm out}$, the 
inclination angle of the stellar spin relative to the outer disc, $\beta_\star=\beta_{\rm out}$, 
evolves according to the equation
\ba
&&J_s{d\over dt}\cos\beta_\star=\bcN\cdot\hatl_{\rm out}-\cos\beta_\star
\,(\bcN\cdot\hat\bomega_s)\nonumber\\
&&\qquad \approx \cN_0\Bigl[\lambda \,(\hatl_{\rm in}\cdot\hatl_{\rm out}
-\cos\beta_\star\,\cos\beta_{\rm in})\nonumber\\
&&\qquad ~~ +n_w\,(\cos\beta_\star-\cos\beta_{\rm in}\,\hatl_{\rm in}\cdot
\hatl_{\rm out}-\cos\beta_\star \sin^2\!\beta_{\rm in})\nonumber\\
&&\qquad ~~ +n_p\,\hat\bomega_s\cdot (\hatl_{\rm in}\times\hatl_{\rm out})
\Bigr].
\label{eq:warpedspinevol}
\ea
Note that this does not depend on the specific form of $\bcN_s$.

For flat discs, equation (\ref{eq:warpedspinevol})
reduces to (Paper I)
\be
\left(\frac{d}{dt} \cos{\beta_\star}\right)_{\rm flat} 
= \frac{\mathcal{N}_0}{J_s} \sin^2{\beta_\star} 
\left(\lambda - \tilde{\zeta} \cos^2\!{\beta_\star} \right),
\label{eq:flatspinevol}
\ee
with
\be
\tilde{\zeta} = \frac{\zeta [ 1 - (r_{\rm in}/r_{\rm int})^3 ]
  \cos^2{\theta_\star}}{6 \eta^{7/2}}.
\ee
In the flat-disc approximation, the star -- disc systems can thus be
divided in two classes with very different long-term spin evolution
(see Fig.~\ref{f1}).  If $\tilde{\zeta} < \lambda$,
$\cos{\beta_\star}$ always increases in time and the system will be
driven towards the aligned state ($\beta_\star=0$).  On the other
hand, if $\tilde{\zeta} > \lambda$, there are two ``equilibrium''
misalignment angles (defined by $d\beta_\star/dt=0$):
\be
\cos{\beta_{\pm}} = \pm \sqrt{\frac{\lambda}{\tilde{\zeta}}}.
\ee
The smaller angle $\beta_+$ corresponds to a stable equilibrium, while
$\beta_-$ is unstable. Thus, the final state of the systems depends on
the initial misalignment angle $\beta_\star(t=0)$. If
$\beta_\star(t=0) < \beta_-$, the system will be driven towards a
moderate misalignment $\beta_+<90^\circ$; otherwise it will evolve
towards a completely anti-aligned configuration ($\beta_\star=180^\circ$).  From 
these results, we can see that, according to the flat-disc approximation, if $\lambda \ll 1$ a misaligned configuration is
strongly favored.

The probability distribution of the different cases for astrophysical systems will thus depend on the unknown value of the parameters of our model, as well as on $\beta_\star(t=0)$ --- which depends
on the formation history of the star -- disc system and is quite uncertain~\citep{blp}. For example, for an isotropic distribution of $\hatl_{\rm out}$ on the unit sphere and $\tilde{\zeta} > \lambda$, a fraction $0.5(1-\sqrt{\frac{\lambda}{\tilde \zeta}})$ of the systems would be anti-aligned, while the rest would tend towards a misalignment $\beta_+$. The real
distribution of disc inclination is certainly more complex, as $\lambda$ and $\tilde \zeta$ will vary
from system to system, and the distribution of the initial misalignment $\beta_\star (t=0)$ is probably not isotropic. Additionally, the orientation of the outer disc might vary in time. For more details on the distribution of final inclination angles $\beta_\star$, see
Paper I (Sec. 5), as well as Section~\ref{sec:aa} of this paper, which discusses a process to reach
anti-alignment starting from $\beta_\star (t=0) < \beta_-$.

\begin{figure}
\begin{centering}
\includegraphics[width=7cm]{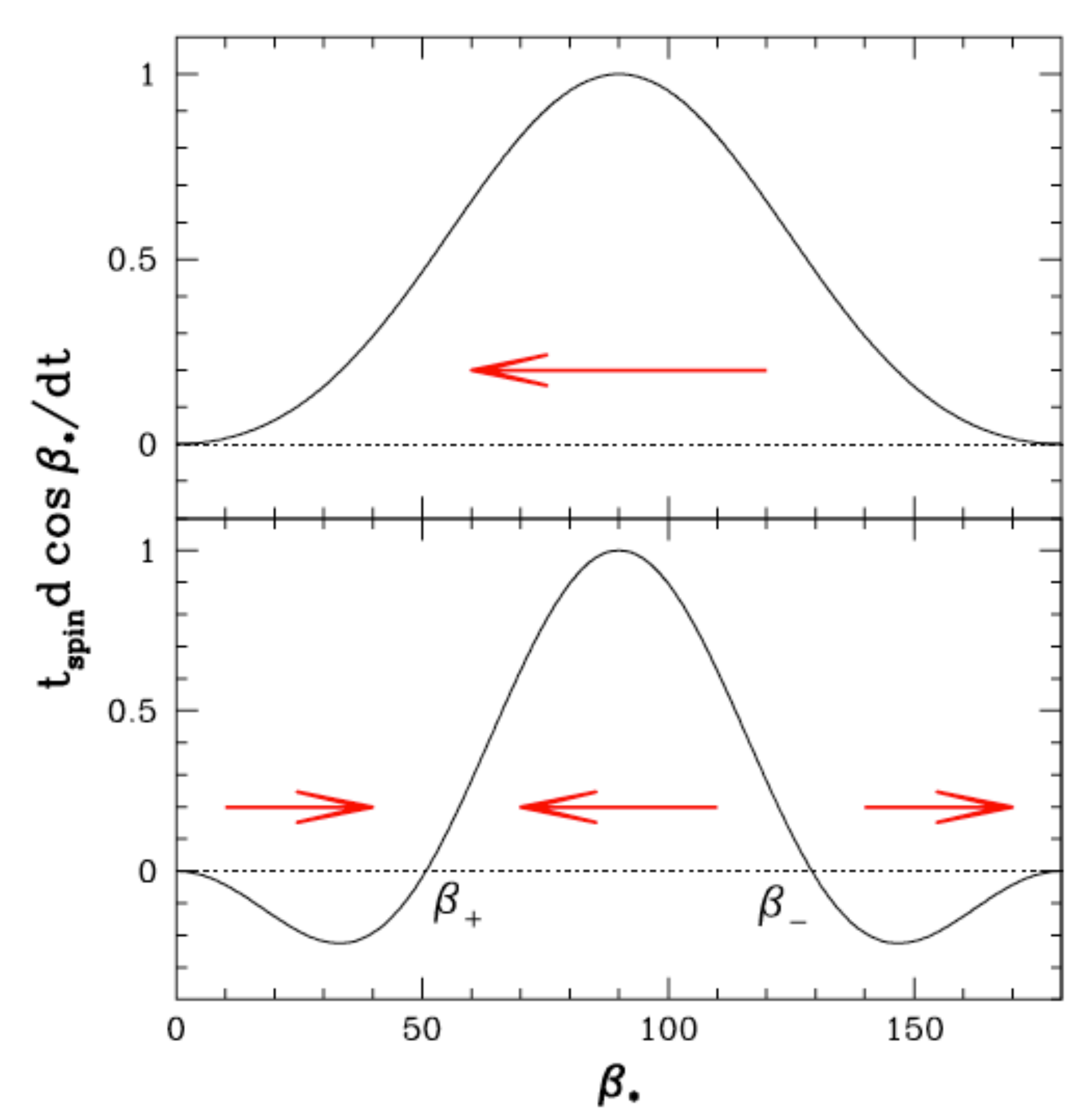}
\caption{Time derivative of the inclination angle of the stellar spin
(in degrees) relative to the outer disc for a flat (unwarped) disc.
The upper panel is for 
$\tilde{\zeta}/\lambda=0.5<1$, in which case
the spin evolves towards alignment. The lower panel is for
$\tilde{\zeta}/\lambda=\sqrt{2}>1$, in which case
the spin either evolves toward $\beta_+\neq 0$ or toward
$\beta_\star=180^\circ$, depending on the initial value of
$\beta_\star$. The quantity $t_{\rm spin}$ is defined in equation~(\ref{eq:tspin}).
The arrows show the direction of the evolution of $\beta_\star$
in different regions of the parameter space.}
\label{f1}
\end{centering}
\end{figure}

In general, the magnetic torques induce disc warping so that $\hatl$
depends on $r$, and equation (\ref{eq:warpedspinevol}) must be used to
determine the long-term spin evolution. Since the disc warp evolution
timescale is much shorter than the stellar spin evolution timescale,
the steady-state warp profile $\hatl(r)$ must be solved before equation
(\ref{eq:warpedspinevol}) can be applied.  In the following sections,
we will show that for most (but not all) realistic choices of the free
parameters in our model, using equation~(\ref{eq:flatspinevol})
instead of equation~(\ref{eq:warpedspinevol}) does not significantly
change our qualitative description of the long term behavior of the
system. We will measure deviations from the flat-disc approximation
through the parameter $\xi$ defined by
\be
\label{eq:paramangevol}
\frac{d}{dt} \cos{\beta_\star} = \xi  
\left(\frac{d}{dt} \cos{\beta_\star}\right)_{\rm flat},
\ee
where the left-hand side is computed using the first line of
equation~(\ref{eq:warpedspinevol}).

\section{Description of Warped Discs: Theory}

As noted in Section 1, the evolution of the coupled star -- disc
system occurs over two different timescales. The first, $t_{\rm disc}$,
characterizes the evolution of the disc under the magnetic torques and
internal stresses towards a warped steady-state configuration,
assuming that the spin of the star $\hat{\bomega}_s$ is fixed. The second,
$t_{\rm spin}$, determines the evolution of $\hat{\bomega}_s$ due to the
combined effects of mass accretion and magnetic torques.  Since we
expect $t_{\rm disc} \ll t_{\rm spin}$, if the orientation of the
outer disc is fixed we can consider that, at all times, the disc is in
a steady-state $\hatl_{\rm eq}(r;\hat\bomega_s)$.  The evolution of
the system is then described by a sequence of steady-state profiles
$\hatl_{\rm eq}(r,\hat\bomega_s(t))$ where $\hat\bomega_s(t)$ evolves
according to equation (\ref{eq:warpedspinevol}) applied to $\hatl(r)=
\hatl_{\rm eq}(r;\hat\bomega_s(t))$.
As discussed before, the disc itself will always show variations
on shorter timescales (of the order of the stellar rotation period), which 
do not affect the secular evolution of the stellar spin and are averaged over in our
description of the system.


Here we describe our method to calculate the evolution and
steady-state of warped discs.

Systematic theoretical study on warped discs began with the work of
\citet{pp1} and \citet{pl2}, who showed
that there are two dynamical regimes for warp propagation in linear
theory (for sufficiently small warps). For high viscosity Keplerian
discs with $\alpha\go \delta\equiv H/r$ (where $\alpha$ is the Shakura-Sunyaev
parameter so that the viscosity is $\nu=\alpha H^2\Omega$), the warp
satisfies a diffusion-type equation with diffusion coefficient
$\nu_2=\nu/(2\alpha^2)$. For low-viscosity discs, on the other hand,
the warp satisfies a wave-like equation and propagates with speed
$\Omega H/2$. In the diffusive regime, the linear theory 
of \citet{pp1} was generalized to large inclination angles by \citet{pr1}
in the limit of small local variations of the disc inclination.
A fully nonlinear theory was derived by \citet{og1}, 
with prescriptions for arbitrary variations of the inclination.
The basic features of the theory
were recently confirmed by the numerical simulations of
\citet{lp1}. For low-viscosity Keplerian 
discs ($\alpha\lo H/r$), the linearized
equations for long wavelength bending waves were derived by
\citet{lo1} and \citet{lop1}, and a theory for
non-linear bending waves was developed by \citet{og2}.


For protostellar discs, recent work by \citet{ter1} suggests that
far away from the star the disc could have a very small 
viscosity parameter ($\alpha \sim 10^{-2}-10^{-4}$), and would 
thus be described by the formalism of~\citet{lo1}. However, 
close to the star (around a few stellar radii) where magnetic effects 
are most important and the disc warp can develop,
the value of the effective viscosity is unknown. Thus in this paper,
we will study both high-viscosity discs and low-viscosity discs.

\subsection{High-Viscosity Discs}

\subsubsection{Evolution Equations}

For viscous discs satisfying $\alpha\go H/r$, we start from the equations
derived by \citet{og1}. The main evolution equations for the disc are
the conservation of mass
\be
\label{MassCon}
\frac{\partial \Sigma}{\partial t} + \frac{1}{r} \frac{\partial}{\partial r} \left( r\Sigma V_{R}\right) =0
\ee
and angular momentum
\ba
&&\frac{\partial}{\partial t}\left(\Sigma r^2 \Omega \hatl \right) +\frac{1}{r}\frac{\partial}{\partial r}\left(\Sigma V_{R}r^3\Omega \hatl\right)=\frac{1}{r}\frac{\partial}{\partial r}\left( Q_1Ir^2\Omega^2\hatl\right) \nonumber \\
&& + \frac{1}{r}\frac{\partial}{\partial r}\left(Q_2Ir^3\Omega^2\frac{\partial \hatl}{\partial r} + Q_3Ir^3\Omega^2\hatl\times\frac{\partial \hatl}{\partial r}\right) + {\bf N}_{m},
\label{MomCon}
\ea 
where $V_{R}$ is the average radial velocity of the fluid at a given radius. 
The coefficients $Q_{1,2,3}$ characterize the magnitude of the various viscous interactions, while
\be
I = \frac{1}{2\pi}\int_0^{2\pi} {\rm d}\phi \int_{-\infty}^{\infty} \rho z^2 {\rm d}z
\ee
depends on the vertical density profile of the disc. The term $\bN_m=\bN_w+\bN_p$
is the external magnetic torque per unit area.

In general, the viscous coefficients $Q_{1,2,3}$ are functions of the
viscosity parameter $\alpha$, the warp amplitude $\psi^2\equiv
|\partial\hatl/\partial\ln r|^2$, and the
disc rotation law $\Omega$. Their values can be obtained through numerical
integration of a set of coupled ODEs \citep{og1}. 
In the limit $\psi^2 \rightarrow 0$, the viscous
coefficients are given by equations [141-143] of \citet{og1}:
\ba
Q_1 &=& -\frac{3 \alpha}{2} + \frac{1}{16 \alpha}\psi^2 +O(\psi^4) \label{Q1}\\
Q_2 &=& \frac{1}{4\alpha} + O(\psi^2)\\
Q_3 &=& \frac{3}{8} + O(\psi^2).
\ea
For $\psi^2=0$ and $Q_3=0$, this is equivalent to the formalism of
\citet{pr1}: the viscosities $\nu_1=\nu$ and $\nu_2$ used by \citet{pr1},
which correspond respectively to the shear viscosity usually
associated with flat discs and the viscous torque working against the
warping of the disc, are proportional to $Q_1$ and $Q_2$. The
additional term $Q_3$ was discovered by \citet{og1}, and contributes
to the precession of a warped disc. For the disc configurations considered
in this paper, the effects of finite $\psi^2$ are small --- hence, our
numerical results will be computed in the limit $\psi^2\ll 1$. However,
we do consider the effects of non-zero $Q_3$.

To obtain numerical solutions to equations
(\ref{MassCon})-(\ref{MomCon}), it is convenient to switch to the
logarithmic coordinate $\rho=\ln{(r/r_{\rm in})}$. We then define the
logarithmic derivative $'=\partial/\partial \ln{r}=\partial/\partial
\rho$ and the warp amplitude $\psi^2=|\hatl'|^2$.
From equations (\ref{MassCon}-\ref{MomCon}), we can derive the
radial velocity
\be
\label{VR}
V_R = \frac{1}{\Sigma r (r^2 \Omega)'} \left[ (IQ_1r^2\Omega^2)'-IQ_2r^2\Omega^2 \psi^2\right].
\ee
Using equation (\ref{VR}) in (\ref{MassCon})-(\ref{MomCon}) then yields
\be
r^2 \frac{\partial \Sigma}{\partial t} = \left[\frac{(IQ_1r^2\Omega^2)'-IQ_2r^2\Omega^2\psi^2}{(r^2\Omega)'}\right]'
\ee
and 
\ba
&&r^2\frac{\partial}{\partial t}\left(\Sigma r^2 \Omega \hatl \right)+
\Bigl[ \frac{r^2\Omega}{(r^2\Omega)'} \left( (IQ_1r^2\Omega^2)'-IQ_2r^2\Omega^2 \psi^2\right) \hatl\nonumber \\
&&-Ir^2\Omega^2\left(Q_1\hatl + Q_2\hatl' + Q_3\hatl\times \hatl'\right)
\Bigr]' = r^2 {\bf N}_{m}.
\ea

\subsubsection{Disc Model}

For our numerical calculations, we consider Keplerian discs. 
If we compare
the projection of (\ref{MomCon}) along $\hatl$ for $\hatl'=0$ with the
standard flat-disc equation
\be
\frac{\partial}{\partial t} (\Sigma r^2 \Omega) + \frac{1}{r}\frac{\partial}{\partial r}\left(\Sigma V_R r^3 \Omega - \nu \Sigma r^3 \frac{\partial \Omega}{\partial r} \right)=0,
\ee
we see that 
\be
Q_1[\psi^2=0] I r^2 \Omega^2 = \nu \Sigma r^3  \frac{\partial \Omega}{\partial r}.
\ee
Using $Q_1[\psi^2=0]=-3\alpha/2$ and $\Omega = \sqrt{GMr^{-3}}$, we then have
\be
I = \Sigma H^2.
\ee
We also rescale the time coordinate by the viscous time evaluated
on the inner edge of the disc: $\tau=t/t_{\rm vis}(r_{\rm in})$,
where 
\be
\label{eq:tvis}
t_{\rm vis}={r^2\over \nu_2},
\ee
and
\be
\nu_2=2Q_2 H^2\Omega\simeq {1\over 2\alpha}H^2\Omega
\ee
is the viscosity associated with the vertical shear in the disc. By
projecting the evolution equation of the disc angular momentum 
onto directions parallel and orthogonal to $\hatl$, we find
\ba
\label{eq:dtlambda}
\frac{\partial}{\partial \tau}\sigma &=& -\rho^{-3/2} \frac{Q_1}{Q_2} \left( 
{\bf S'}\cdot\hatl \right) \\
\label{eq:dtl}
\frac{\partial}{\partial \tau} \hatl &=& - \rho^{-3/2} \frac{Q_1}{\sigma Q_2}
\bigg[ \left[{\bf S'}-({\bf S'}\cdot\hatl)\hatl\right]+ \\
&&\frac{(\hat\bomega_s \cdot \hatl) }{\rho^{3}\eta^{7/2}} 
\left( \frac{F(\theta_\star)}{\pi D(\rho)}
\hat{\bomega}_s \times \hatl - \frac{\zeta \cos^2{\theta_\star}}{4} \hatl \times ( \hat{\bomega}_s 
\times \hatl )
\right)\bigg]
\nonumber
\ea
where the new variables $\sigma$, ${\bf S}$ and $\rho$ are defined by
\ba
\sigma &=& \frac{\Sigma r}{(r \Sigma_{\rm flat})|_{r=r_{\rm in}}}\\
\label{eq:defS}
{\bf S} &=& \left( \sigma' - \frac{\sigma}{2} - \frac{Q_2}{Q_1}\sigma \psi^2 \right) \hatl
-\frac{Q_2}{2Q_1}\sigma \hatl' - \frac{Q_3}{2Q_1} \sigma \hatl \times \hatl'\\
\rho &=& \frac{r}{r_{\rm in}}.
\ea
and $\Sigma_{\rm flat}$ is the surface density of a flat disc
\be
\label{eq:sflat}
\Sigma_{\rm flat}=\frac{\dot{M}}{3\pi \nu}
=\frac{\dot{M}}{3\pi \alpha H^2\Omega}.
\ee
Equations (\ref{eq:dtlambda}) and (\ref{eq:dtl}) form our model for 
the evolution of viscous discs interacting with a magnetic star. Note
that as $\hatl$ is a unit vector, it only corresponds to two degrees of freedom in the
system. Accordingly, equation~(\ref{eq:dtl}) guarantees that $\partial \hatl / \partial \tau$ is orthogonal to $\hatl$. In practice, to avoid introducing a preferred direction in the system (as we want to allow arbitrary inclination angles for the disc), we evolve all 3 components of $\hatl$, but
normalize $\hatl$ at each timestep to the accumulation of numerical errors.

\subsubsection{Steady-State Equations}
\label{sec:warpsse}

From equations (\ref{eq:dtlambda}), (\ref{eq:dtl}) and
(\ref{eq:defS}), it is fairly easy to derive the equations defining
the steady-state configuration of the disc. If we set 
$\partial \hatl/ \partial \tau = 0$ and $\partial \sigma / \partial \tau = 0 
= {\bf S}'\cdot\hatl$ in (\ref{eq:dtl}), we obtain
\be
{\bf S}' =  \frac{(\hat\bomega_s . \hatl) }{\rho^{3}\eta^{7/2}} 
\left(  \frac{\zeta \cos^2{\theta_\star}}{4} \hatl \times( \hat{\bomega}_s 
\times \hatl )-\frac{F(\theta_\star)}{\pi D(\rho)}
\hat{\bomega}_s \times \hatl \right).
\ee
Equation (\ref{eq:defS}) projected onto $\hatl$ gives
\be
\label{eq:sslambda}
\sigma' = \sigma \left(\frac{1}{2} + \frac{Q_2}{Q_1} \psi^2 \right)
 + {\bf S}\cdot\hatl,
\ee
and projected in the plane orthogonal to $\hatl$ gives
\be
\label{eq:ssl}
\hatl' = \frac{2Q_1}{Q_2 \sigma}\left[({\bf S.}\hatl)\hatl - {\bf S}\right] - \frac{Q_3}{Q_2} (\hatl \times \hatl').
\ee
For $Q_3=0$, we thus have a set of first order differential equations of the form
${\bf U}' = {\bf F}({\bf U})$. Given appropriate boundary conditions at $r_{\rm in}$, it can easily be
solved by numerical integration. For $Q_3\neq 0$, we can still perform numerical integration if we
consider (\ref{eq:ssl}) as an implicit equation for $\hatl'$ which has to be solved at each step of the
integration algorithm.

In practice however, the boundary conditions are imposed partly at the inner edge $r_{\rm in}$ and
partly at the outer edge $r_{\rm out}$. Indeed, we consider the orientation of the outer disc to be
fixed
\be
\hatl(r_{\rm out}) = \hatl_{\rm out}
\ee
and the mass accretion rate to be constant
\be
\label{Mdot0}
\dot{M} = -2\pi r V_R \Sigma
\ee
(the sign is chosen so that $\dot M > 0$ for $V_R<0$). We also impose
a zero-torque boundary condition at the inner edge
\be
\hatl'(r_{\rm in}) = 0
\ee
and set the surface density there to
\be
\label{SBC}
\Sigma(r_{\rm in}) = \sigma_{\rm in} \Sigma_{\rm flat}(r_{\rm in})
\ee
for some freely specifiable scalar $\sigma_{\rm in}$. Combining
(\ref{Mdot0}) with the zero-torque boundary condition and equations
(\ref{VR}) and (\ref{eq:sflat}), we obtain a simple boundary condition on
$\sigma'$ at $r_{\rm in}$:
\be
\sigma'[r_{\rm in}]=\frac{1}{2},
\ee
while (\ref{SBC}) gives the value of $\sigma[r_{\rm in}]$:
\be
\sigma[r_{\rm in}] = \sigma_{\rm in}.
\ee
We thus have 4 boundary conditions at $r_{\rm in}$ (on $\hatl'$,
$\sigma$ and $\sigma'$) and 2 at $r_{\rm out}$ (on $\hatl$). To
solve the system numerically we use a shooting method starting at
$r_{\rm in}$. Writing $\hatl = (\cos{\beta} \cos{\gamma}, \cos{\beta}
\sin{\gamma}, \sin{\beta})$ and $\hat{\bomega}_s=(0,0,1)$, we use a
2-D Newton-Raphson method to solve for the values of $\beta[r_{\rm
    in}]$ and $\gamma[r_{\rm in}]$ leading to a solution satisfying
$\hatl (r_{\rm out})=\hatl_{\rm out}$.  The system of first-order ODEs
which has to be solved at each iteration of the Newton-Raphson
algorithm is treated using the 5th order {\it StepperDopr5} method of
\citet{NR}, and the integration is performed under the constraint
$|\hatl|=1$.

\subsection{Low-Viscosity discs}

\subsubsection{Evolution Equations}
\label{sec:evollv}

For discs with a viscosity parameter small compared to the thickness
($\alpha\lo \delta=H/r$), we can no longer use the evolution
equations of \citet{og1}.  In this case, disc warps 
propagate as bending waves. In the linear regime, 
the warp evolution equations were derived by \citet{lo1}:
\ba
\label{eq:dtlvl}
&&\Sigma r^2 \Omega \frac{\partial \hatl}{\partial t} = \frac{1}{r} \frac{\partial {\bf G}}{\partial r} + {\bf N}_m, \\
\label{eq:dtG}
&&\frac{\partial {\bf G}}{\partial t} = \left(\!\frac{\Omega^2 - \Omega_r^2}
{2\Omega}\!\right) \hatl \times {\bf G} - \alpha \Omega {\bf G} 
+ \frac{\Sigma r^3c_s^2 \Omega}{4} 
\frac{\partial \hatl}{\partial r},
\ea
where $c_s=H\Omega_z$ is the disc sound speed,
$\Omega_r$ and $\Omega_z$ are the radial epicyclic frequency and
the vertical oscillation frequency associated with circular orbits at
a given radius from the star, ${\bf G}$ is the internal torque of the
disc, and $\Sigma =\Sigma_{\rm flat}$ is the surface density. 
These equations are only valid for
$\alpha\lo \delta$, $|\Omega_r^2-\Omega^2|<\delta \Omega^2$ and
$|\Omega_z^2-\Omega^2|<\delta \Omega^2$. In the following, we shall use
$\Omega_r=\Omega_z=\Omega$, although we verified that small deviations
from these equalities do not significantly modify our results.

Equations (\ref{eq:dtlvl})-(\ref{eq:dtG}) admit wave solutions. 
We define a Cartesian coordinate system so that 
$\hat l_z \simeq 1$ and $|\hat l_{x,y}|\ll 1$, and 
the internal torque ${\bf G}$ acts in the $xy$-plane.
Consider a local (WKB) wave with $\hatl_{x,y}, {\bf G}\propto
e^{ikr-i\omega t}$ in a Keplerian disc
with $\bN_m=0$. For $\omega\ll \Omega$, the dispersion
relation of the wave is, neglecting the damping term $\alpha \Omega {\bf G}$
in~(\ref{eq:dtG}),
\be
{\omega\over k}=\pm {c_s\over 2}=\pm \frac{H\Omega}{2},
\ee
with the eigenmodes satisfying 
\be
\hatl_{x,y}=(\hatl_{x,y})_\pm
\equiv \mp {2\over r^3 c_s\Omega\Sigma}{\bf G}_{x,y}
=\mp {6\pi\alpha \delta\over \dot M r^2\Omega}{\bf G}_{x,y}.
\ee
The $+$ mode and $-$ mode correspond to the outgoing and ingoing bending waves,
respectively.

A generic warp perturbation will not behave as pure eigenmodes.
For numerical evolutions, it is convenient to define the variables
\be
{\bf V}_{\pm x,y} = \hatl_{x,y} \mp \sqrt{\frac{r_{\rm in}}
{r}} \frac{6\pi \alpha \delta}{\dot{M} r_{\rm in}^2
\Omega[r_{\rm in}]} {\bf G}_{x,y}.
\ee
Then, the evolution equations for the disc can be written as
\ba
\label{eq:Vwave}
&&\frac{\partial}{\partial \tau}{\bf V}_{\pm} = \frac{1}{2\rho^{3/2}}
\bigg[ \mp {\bf V}'_{\pm} 
+ ({\bf V}_- - {\bf V}_+)
\left( \frac{1}{4} \pm \frac{\alpha}{\delta} \right) 
 \bigg] \\
 \nonumber
 && + \frac{\cos{\beta}}{\rho^5 \eta^{7/2}} 3\alpha \delta \left[ \frac{F(\theta_\star)}{\pi D(\rho)}
 \hat{\bomega}_s \times \hatl - \frac{\zeta \cos^2{\theta_\star}}{4} \hatl \times (\hat{\bomega}_s 
 \times \hatl) \right].
\ea
Here the dimensionless time $\tau = t \delta \Omega(r_{\rm in})$ and
length $\rho = r/r_{\rm in}$ are chosen so that the sound speed
at the inner edge of the disc is $c_s(r_{\rm in})=H\Omega_z=1$.
For the computation of the magnetic torque, we use the following
approximations, accurate to first order in $l_{x,y}$:
\ba
\hat{\bomega}_s \times \hatl &=& -\cos{\beta}\hat{l}_y \hat{e}_x+(\sin{\beta} + \cos{\beta} \hat{l}_x) \hat{e}_y\\
\hatl \times (\hat{\bomega}_s \times \hatl) &=& -(\sin{\beta}+\cos{\beta} \hat{l}_x )\hat{e}_x 
-\cos{\beta}\hat{l}_y \hat{e}_y.
\ea

The boundary conditions are particularly simple to implement for this
choice of variables. At the outer edge of the disc, we require the
ingoing mode to vanish
\be
{\bf V}_-(r_{\rm out})=0.
\ee
At the inner edge, we impose the zero-torque boundary
condition $\hatl'=0$, ${\bf G}=0$, which can be written in terms of our
evolution variables as
\ba
{\bf V}_-(r_{\rm in})&=&{\bf V}_+(r_{\rm in}),\\
{\bf V}_-'(r_{\rm in})&=&-{\bf V}_+'(r_{\rm in}).
\ea
In terms of the propagation of bending waves, this corresponds to the 
requirement that the waves be reflected at the inner edge of the disc.

\subsubsection{Steady-State Warp}
\label{sec:sslv}

The steady-state profile of low-viscosity discs can be obtained by
numerical integration of equation (\ref{eq:Vwave}) or equations
(\ref{eq:dtlvl})-(\ref{eq:dtG}), by setting $\partial / \partial \tau
= 0$. In practice however, the steady-state profile of a low-viscosity
disc is nearly always very well approximated by a flat disc
profile. The amount of disc warping can then be 
evaluated analytically. Noting 
that $\Sigma r^3 c_s^2\propto r^{3/2}$, equations
(\ref{eq:dtlvl})-(\ref{eq:dtG}) can be combined to give
\be
\frac{\partial}{\partial r} \left(\rho^{3/2} \frac{\partial}{\partial r} \hatl \right)
\simeq -\frac{4\alpha r {\bf N}_m}{(r^3\Sigma c_s^2)_{\rm in}}.
\ee
Since ${\bf N}_m$ is falling rapidly with $r$ (${\bf N}_m \sim r^{-5}$), 
and $\partial\hatl/\partial r=0$ at $r=\rin$, we integrate the above equation 
from $\rin$ to $r$: 
\be
\frac{\partial}{\partial r} \hatl \simeq  
{4\alpha\over 3} {(r^2\bN_m)-(r^2\bN_m)_{\rm in}
\over \rho^{3/2}(r^3\Sigma c_s^2)_{\rm in}}.
\label{eq:lindl}
\ee
Integrating from $r_{\rm out}$ to $\rin$, we then obtain
\be
\hatl_{\rm in}-\hatl_{\rm out}\simeq \left({16\alpha \bN_m\over 7\Sigma c_s^2}
\right)_{\rm in}.
\label{eq:deltal}
\ee
Using Eqs.~(\ref{eq:torquew})-(\ref{eq:torque}), we have
\be
|\hatl_{\rm in}-\hatl_{\rm out}|\simeq 
{4\over 7}\Bigl[t_{\rm vis}(|\Gamma_w|+|\Omega_p|)\sin(2\beta)\Bigr]_{\rm in},
\ee
where $t_{\rm vis}=r_{\rm in}^2/\nu_2$ is the viscous timescale for the warp,
with $\nu_2=c_sH/(2\alpha)$.  Thus, the distortion of the disk can be
seen as arising from the warping and precessional torques acting over
the disc during a time of order the viscous time scale at the inner
disc edge (where the magnetic torques are the strongest).
Projecting Eq.~(\ref{eq:deltal}) in the direction of the stellar spin axis
$\hat\bomega_s$ and using Eqs.~(\ref{eq:torquew})-(\ref{eq:torque}), we have
\be
\cos\beta_{\rm in}-\cos\beta_{\rm out}=-{8\over 7}\left(t_{\rm vis}\Gamma_w
\cos\beta\sin^2\beta\right)_{\rm in}.
\label{eq:cosbeta}\ee
Since
\be
\label{eq:tvisgam}
\left(t_{\rm vis}\Gamma_w\right)_{\rm in}={3\alpha^2\zeta\over 2\eta^{7/2}}
\cos^2\theta_\star,
\ee
we see that as long as $\alpha^2\zeta\ll\eta^{7/2}$, a condition satisfied for 
most parameters, the warp across the whole disc is small:
\be
|\beta_{\rm in}-\beta_{\rm out}|\simeq
\frac{6\alpha^2\zeta \sin{(2\beta)}\cos^2{\theta_\star}}{7\eta^{7/2}}.
\label{eq:approxwarp}
\ee
For example, with $\eta\go 0.5$, we find that for all discs 
$|\beta_{\rm in}-\beta_{\rm out}|\ll 1$ if $\alpha\lo 0.15$. 
This is almost certainly true for discs in which bending waves can propagate.

It is important to note that, although the approximate analytical
expression of the global disc distortion derived above is based on
low-viscosity discs, our result for $|\beta_{\rm in}-\beta_{\rm out}|$
is also valid for higher-viscosity discs.  Indeed, in the linear
regime and for Keplerian discs, the steady-state equations are
identical regardless of the viscosity regime considered.

\section{Steady-State Profile of Warped Discs and Back-Reaction on Stellar Spins}

Using the numerical scheme presented in Section \ref{sec:warpsse}, we can now determine
the time-averaged steady-state profile of the disc under the influence of the torques exerted by a magnetic star.
The characteristics of the warped disc will of course vary with the choice of the free parameters
included in our theoretical model. We begin our study by showing results for two standard
discs, chosen so that they belong to the two classes of long term stellar spin evolution predicted in
Section \ref{sec:spinevol} when the accretion parameter defined in 
equation~(\ref{eq:Nl}) is $\lambda \approx 0.5$ (a typical value in the allowed range
$0 \le \lambda \le 1$). 
We then vary the disc parameters, and discuss their
influence on the disc profile, and on the spin evolution. 
Finally, we check that, as predicted in section~\ref{sec:sslv}, low-viscosity discs,
which follow the different evolution equations described in section~\ref{sec:evollv} 
(valid for $\alpha \lo \delta=H/R$) 
have only negligible steady-state warps and are for all practical purposes well described by the flat-disc
approximation.

Our base models are discs with viscosity $\alpha=0.15$ and thickness $\delta=0.1$. We fix the 
surface density at the inner boundary by setting $\sigma_{\rm in}=1.0$ so that it is equal to the surface density of a flat disc~(\ref{eq:sflat}), choose the inclination 
angle of the outer disc $\beta_{\rm out}=\beta_\star=10^{\circ}$ and the magnetic inclination angle $\theta_
\star=30^{\circ}$ with respect to the spin $\hat\bomega_s$. The star is assumed to have mass $M_
\star= M_{\odot}$ and radius $R_\star = 2 R_{\odot}$ The strength of the magnetic field is chosen 
so that $r_{\rm in}=2.5 R_*$ (Corresponding to $B_\star\sim 1kG$ for typical parameters, see
Eq.~[\ref{alfven}]), and the action of the torque ${\bN}_m$ is limited to the region $r_{\rm in} \leq r \leq r_{\rm int}=1.5r_{\rm in}$. 
The accretion rate is $\dot{M}=10^{-8} M_{\odot}/{\rm yr}$, and we put 
the outer disc boundary at $r_{\rm out} = 10^4 r_{\rm in}$ (corresponding to $r_{\rm out} \approx 
250 {\rm AU}$, a size typical of the observed protoplanetary discs). This disc has small values of $\psi^2$ 
everywhere, and accordingly we neglect the nonlinear terms in $Q_i$ (but we keep $Q_3=3/8$). 
The other parameters are chosen to be $\zeta=1$, $f=0$, and either $\eta=1$ 
(so that the long-term evolution of the system aligns the spin axis $\hat\bomega_s$ with the disc
axis) or $\eta=0.5$ (for 
which the flat-disc approximation predicts a long term misalignment toward 
$\beta_+ \approx 45^{\circ}$ if 
the initial disc has $\beta_\star \leq 135^\circ$). 

We should note that these parameters are purposefully chosen to test the limits of
the flat disc approximation. Our choices $M_\star$, $R_\star$, $B_\star$, $\delta$ and 
$M_\odot$ are relatively standard values for protoplanetary discs around T-Tauri stars 
[see~\cite{bourev} and references therein],
while there are no particular reasons to prefer any specific orientation of the magnetic dipole 
$\theta_\star$. But $\alpha=0.15$ is larger that recent estimates of the viscosity in the outer parts of 
the disc~\citep{ter1}, and probably on the high end of what can be expected in the inner disc. 
However, we have shown that smaller values of $\alpha$ lead to smaller amplitudes of the 
steady-state warp (the warp amplitude is proportional to $\alpha^2$).
Thus, the flat disc approximation is more likely to be satisfied at low
viscosities.

In order to analyze the radial variations of the disc warp profile, we define the tilt $\beta[r]$ and the twist
$\gamma[r]$ by 
\be
\hatl[r]=(\sin\beta[r] \cos\gamma[r], \sin\beta[r] \sin\gamma[r],\cos\beta[r]),
\ee
with
the convention that $\gamma[r_{\rm out}]=0$. Some parameters of the system can be varied without modifying the dimensionless solution for the profile of the surface density 
$\sigma(\rho)$  and the orientation of the disc $\hatl(\rho)$: modifications of $\dot M$, $M_\star$, $R_\star$ or $r_{\rm in}$ 
(at constant $\eta$, $r_{\rm out}/r_{\rm in}$ and $r_{\rm int}/r_{\rm in}$) will influence the values of the timescales $t_{\rm vis}$ and $t_{\rm spin}$, but not $\beta[\rho]$ or
$\gamma[\rho]$. Thus, the steady-state profile can be solved while keeping these parameters fixed
without any loss of generality. The disc profile in physical units [$\Sigma(r)$, $\hatl(r)$]
can easily be retrieved from the dimensionless solution [$\sigma(\rho)$, $\hatl(\rho)$].
Additionally, the four parameters $(\eta,\theta_\star,f,\zeta)$
correspond to only two degrees of freedom in the model, through the quantities
\ba
c_1 &=&{ \zeta \cos^2{\theta_\star} \over \eta^{3.5}} \\
c_2 &=& {F(\theta_\star) \over \eta^{3.5}} = {2f \cos^2{\theta_\star} - \sin^2{\theta_\star}
\over \eta^{3.5}}.
\ea
We will thus limit ourselves to variations of $\zeta$ and
$f$. Varying the thickness $\delta$ of the disc has very similar effects: it changes the value of the
function D(r) at small radii, effectively modifying the value of $c_2$ close to $r_{\rm in}$. As the
magnetic torques mostly affect the region close to the inner edge of the disc, the influence of 
$\delta$ is similar to that of $F(\theta_\star)$. 
Finally, we are also free to modify the boundary conditions used, and in particular the choices
of $r_{\rm out}$ and $\sigma_{\rm in}$. Varying $r_{\rm out}$ seems to have only negligible effects, as
long as $r_{\rm out}/r_{\rm in}$ is large enough for a steady-state solution to exist. Decreasing
$\sigma_{\rm in}$, on the other hand, leads to more significant changes in the warp profile. 
A small $\sigma_{\rm in}$ favors warping disc, so that a decrease of 
$\sigma_{\rm in}$ has an effect similar to increasing
both $c_1$ and $c_2$.

Thus, the effects of varying various parameters of the system can be examined with our
standard discs, by varying only two parameters, $\zeta$ (or $c_1$) and $f$ (or $c_2$). 
In section~\ref{sec:stddisc}, we present our results for our two standard discs, which are similar except for the value of the parameter $\eta$ (changing $\eta$ correspond to a rescaling of
 both the warping and the precessional torque). Then, in section~\ref{sec:nwinf}, we study 
 variations of the warping torque alone, by modifying the value of the parameter $\zeta$
 characterizing the strength of the toroidal field in the disc. The influence of the precessional
 torque is studied in more details in section~\ref{sec:npinf}, through variations of the parameter $f$
 (related to the ability of the time-varying component of the vertical magnetic field to penetrate
 the disc). Finally, in section~\ref{sec:Q3} we comment on the influence of the
 parameter $Q_3$, which was usually neglected in previous studies of warped discs.

\subsection{Standard disc results}
\label{sec:stddisc}

The profile for the tilt and twist angles of our standard 
configurations (Fig. \ref{fig:tiltstd}) show a relatively weak warping of the disc. For the disc with weaker magnetic 
interactions ($\eta=1$), the difference in tilt  
between the inner and outer edges is about $0.17^{\circ}$ and the twist over the whole disc is 
$1.2^{\circ}$, while for stronger interactions ($\eta=0.5$) the disc is tilted by $1.8^\circ$ and 
twisted over $16^\circ$. 
These warps are comparable in magnitude to what we could have predicted using the 
approximate equations~(\ref{eq:deltal})
and~(\ref{eq:approxwarp}). In particular, formula~(\ref{eq:approxwarp}) applied to these two 
choices of parameters predicts tilts of $0.28^\circ$ and $3.2^\circ$, respectively,
with most of the difference between the approximate formula and the numerical results due to the cutoff applied to the magnetic torques at $r=r_{\rm int}$, neglected in the derivation
of~(\ref{eq:approxwarp}). 

Note that we choose to vary the parameter $\eta$ defined in equation~(\ref{alfven}) 
as it conveniently modifies the effective strength of both magnetic torques in our model.
In practice, $\eta$ is determined by the 
geometry of the accretion flow, while unknown physical parameters such as the dipole strength 
$\mu$, its orientation $\theta_\star$, the surface density at the inner edge of the disc 
$\sigma_{\rm in}$ or the magnetic twist parameter $\zeta$ will vary from
system to system.

 Given the small warp, the evolution of the misalignment angle $\beta_\star$ 
 between $\hat{\bomega}_s$ and ${\hatl}_{\rm out}$ is well approximated 
 by equation~(\ref{eq:warpedspinevol}) with ${\hatl}_{\rm in} = {\hatl}_{\rm out}$: if we
compute $\bcN$ from the steady-state profile $\hatl(r)$, we find that the parameter $\xi$ in
equation (\ref{eq:paramangevol}), which parametrizes deviations from the flat
disc approximation ($\xi=1$ for a flat disc) is $\xi = 0.997$ for $\eta=1$ and 
$\xi = 0.86$ for $\eta=0.5$, if we set the accretion parameter $\lambda$ to 0 
(we choose $\lambda=0$ when computing $\xi$ in order to measure directly differences in the
effect of the back-reaction magnetic torques between the flat-disc model and the warped disc
steady-state, disentangled from the effect of angular momentum
accretion).

However, even a small disc warp can significantly change the critical angles $\beta_\pm$ for which 
$d\beta_\star/dt=0$. By varying $\beta_\star=\beta(r_{\rm out})$, we can determine the values of 
$\beta_\pm$ numerically. For $\eta=0.5$ and $\lambda=0.5$, we find $\beta_+ = 32^\circ$ and 
$\beta_-=148^\circ$ which are quite different from the prediction of the flat-disc approximation
($\beta_+=45^\circ$, $\beta_-=135^\circ$). This is due mainly due to the effect of the twist of the
disc. In the flat-disc approximation, $\gamma=0$ and the back-reaction due to the precession
torque has no effect on the evolution of the stellar spin. But as long as $\gamma_{\rm in}<\pi$, 
that back-reaction will tend to align the stellar spin and the disc orbital angular momentum, and
this effect can be large enough to significantly shift the value of $\beta_\pm$ (see also 
subsection~\ref{sec:npinf}).
The same effect can also modify the qualitative behavior of systems for which the predicted misalignment angle $\beta_+$ is close to $0$, in such a way that the stable configuration at $\beta_+$ no longer exist. The orbital angular momentum of the disc would then
be expected to align with the direction of the stellar spin.

\begin{figure}
\includegraphics[width=8cm]{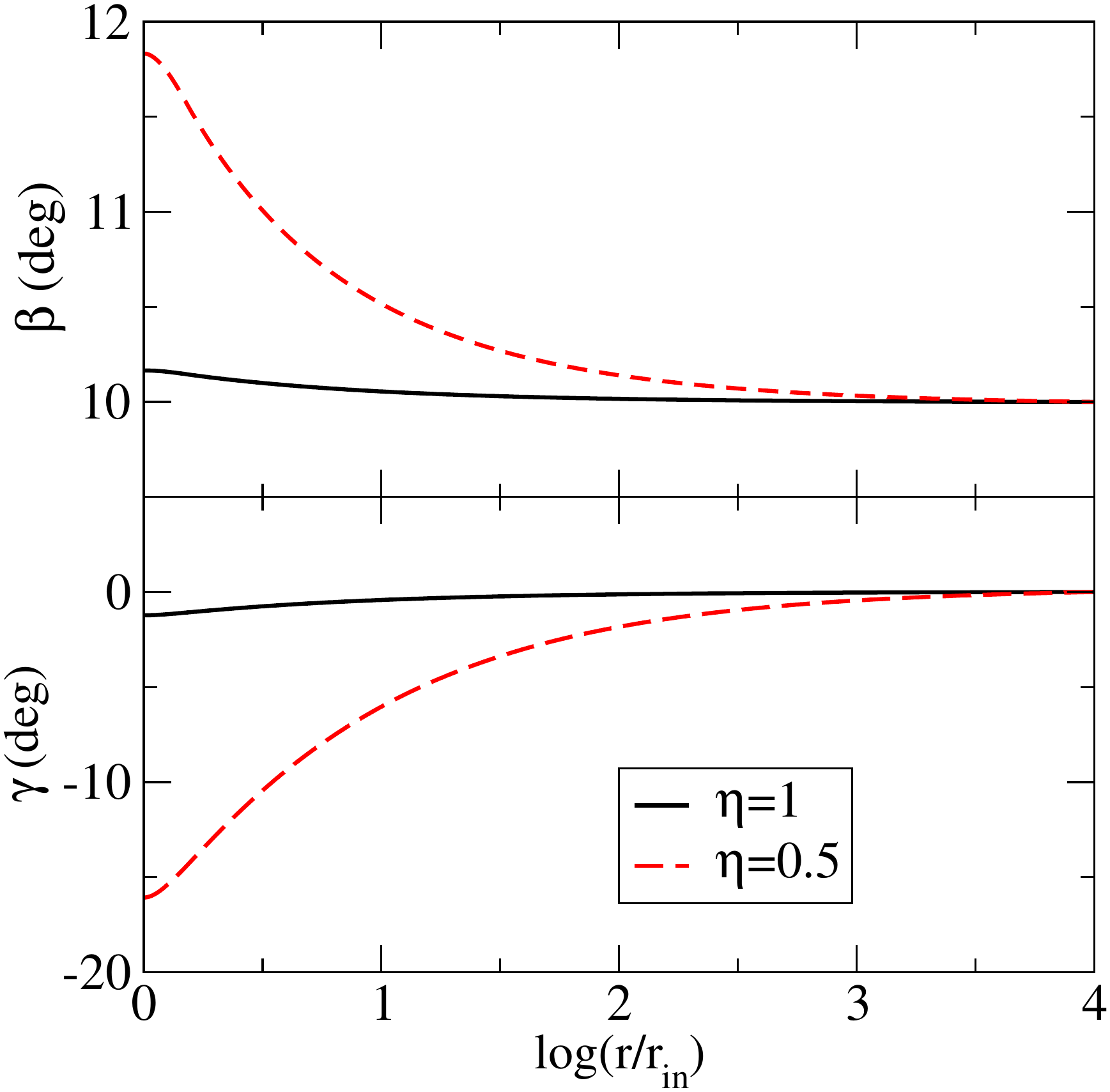}
\caption{{\it Upper panel:} Disc tilt angle $\beta$ for our standard disc, with $\eta=1$ and $\eta=0.5$. {\it Lower panel: } Twist angle $\gamma$ for the same parameters.}
\label{fig:tiltstd}
\end{figure}

\subsection{Large warping torques}
\label{sec:nwinf}

Realistic discs are expected to have parameters $\zeta \sim 1$ (characterizing the azimuthal magnetic twist) and $\eta \sim 0.5$ (characterizing the inner disc radius; see Eq.~[\ref{alfven}]), but the exact values of those 
parameters are unknown (see Section~\ref{sec:analytic}). By increasing $\zeta$ we see that the approximation $
\hatl(r_{\rm in}) = \hatl(r_{\rm out})$ can break down for high-viscosity discs. 
In Fig. \ref{fig:zetaseqtilt}, we show the variation of the steady-state disc profile when $\zeta$ is varied
 between 1 and 5 
for $\eta=0.5$ and our standard disc parameters. Clearly, there can be large 
differences between the orientation of the disc at its inner and outer edges when $\zeta \go 1$. 
In Fig. \ref{fig:xizeta}, we show the value of $\xi$ [Eq.~\ref{eq:paramangevol}]
for various choices of $\zeta$. At low 
$\zeta \lo 4$ and for the choice of accretion parameter $\lambda=0$, 
the flat-disc approximation predicts the magnitude of the back-reaction 
torques acting on the star within a factor of 2. Deviations at low $\zeta \approx 0.5$ are due
to the relatively large influence of the precessional torque (which is independent of $\zeta$)
when the warping torque becomes small.
The long-term evolution of the stellar spin direction will remain similar to the flat-disc
predictions, with a stable configuration at 
some misalignment angle $\beta_+ \neq 0$ for most values of the accretion parameter
$\lambda$. But for $\zeta \go 4.5$, the twist is so large that the 
behavior is the opposite of what would be predicted by our approximate flat-disc formula: 
the back-reaction 
tends to align the disc and the spin of the star. This shows that at large $\zeta$ we must 
determine for each set of parameters the profiles $\beta[r]$ and $\gamma[r]$ in order to predict the long term evolution of the stellar spin. 
\begin{figure}
\includegraphics[width=8cm]{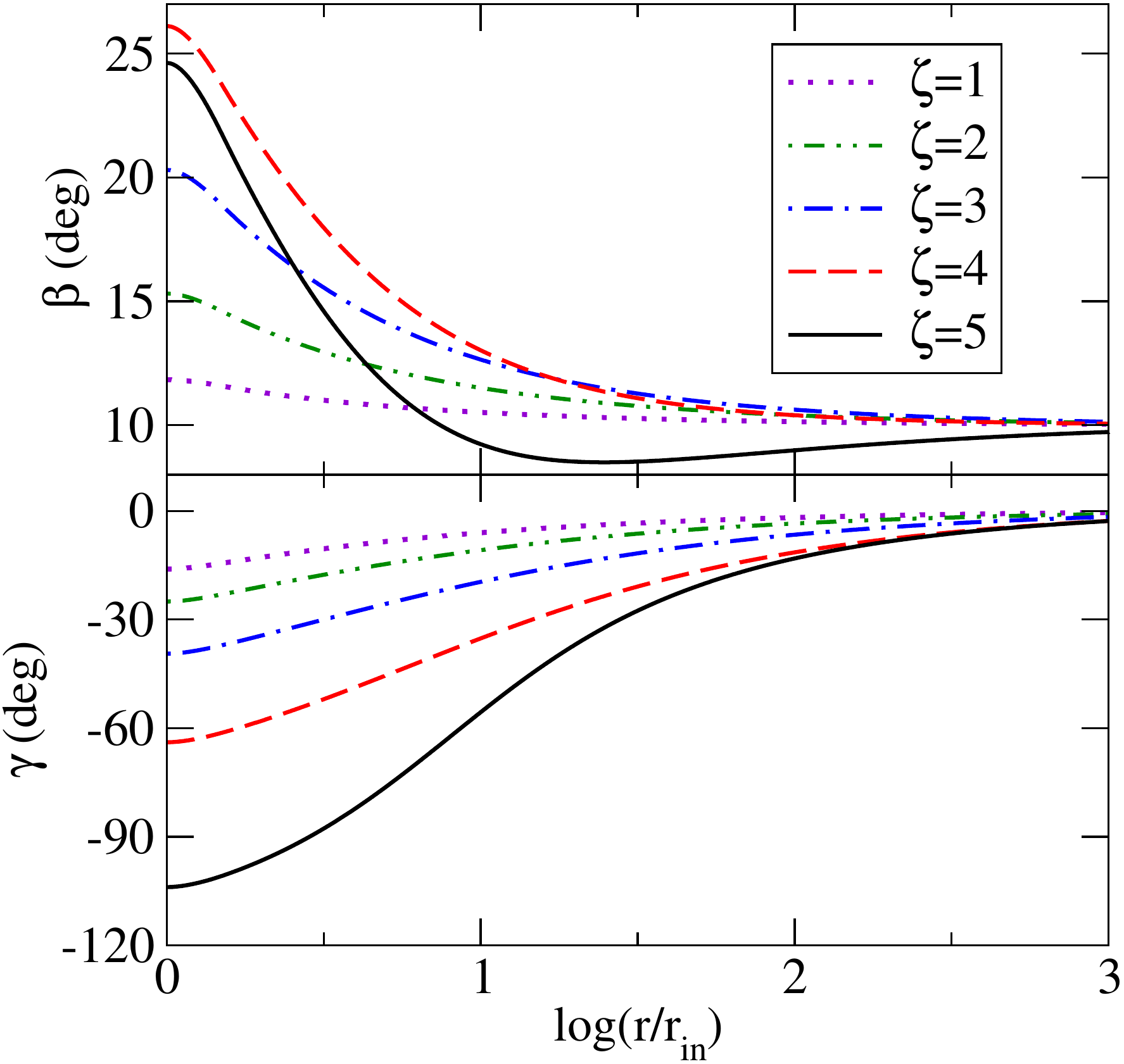}
\caption{{\it Upper panel:} Disc tilt angle $\beta$ for different choices of $\zeta$ values, all with $\eta=0.5$. {\it Lower panel: } Twist angle $\gamma$ for the same disc parameters. The outer edge
of the disc is at $r_{\rm out}=10^4r_{\rm in}$.}
\label{fig:zetaseqtilt}
\end{figure}
\begin{figure}
\includegraphics[width=8cm]{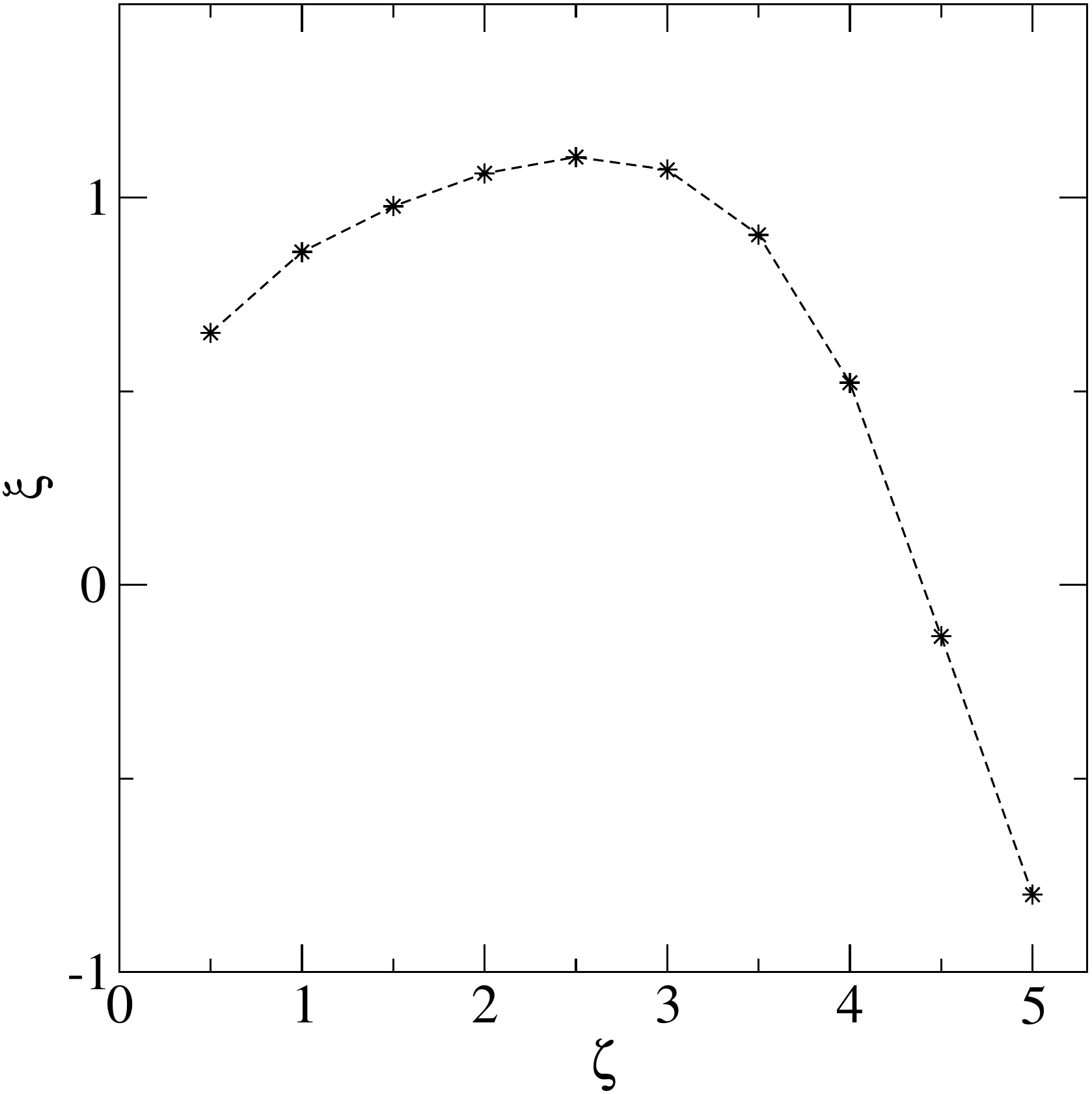}
\caption{Variation of the parameter $\xi$ characterizing the deviations from the flat-disc
 approximation [see Eq.~(\ref{eq:paramangevol})] as a function of $\zeta$, for a sequence
 of discs with $\eta=0.5$ and the choice
$\lambda=0$ for the accretion parameter [see Eq.~(\ref{eq:Nl})].}
\label{fig:xizeta}
\end{figure}
As an example, we construct sequences of steady-state disc configurations for a fixed $\zeta$, 
varying the inclination angle of the outer disc $\beta_\star$. In Figs. \ref{fig:zeta1seq}-\ref{fig:zeta5seq}, we show the resulting $d\cos{\beta_\star}/dt$ for $\zeta=1,3$ and $5$, and compare with the predictions of the flat-disc approximation. For $\zeta=1,3$, the general behavior is similar to what the flat-disc approximation predicts. As seen in subsection~\ref{sec:stddisc},
the precessional torque will favor alignment of the stellar spin with the disc orbital angular momentum,
so that the numerical results usually show that $\beta_{+,Num} \leq \beta_{+,Flat}$ --- at least
as long as $F(\theta_\star)$ is of order unity. 
For $\zeta=5$, however, significant differences become visible. At small inclination angles, the system will evolve towards $\beta_\star=0$, while at  large inclinations, the system will evolve towards $\beta_\star \approx 165^\circ$. In the intermediate region 
$15^\circ \lo \beta_\star \lo 135^\circ$, the system will evolve towards 
$\beta_\star \approx 50^\circ$ (for $\lambda=0.5$). 
Finally, for larger $\zeta$ we are in a completely different regime: for some inclinations, two steady-state solutions exist. Clearly, to determine which of those steady-state solution is relevant
requires numerical integration of the time evolution of the star-disc system.
\begin{figure}
\includegraphics[width=8cm]{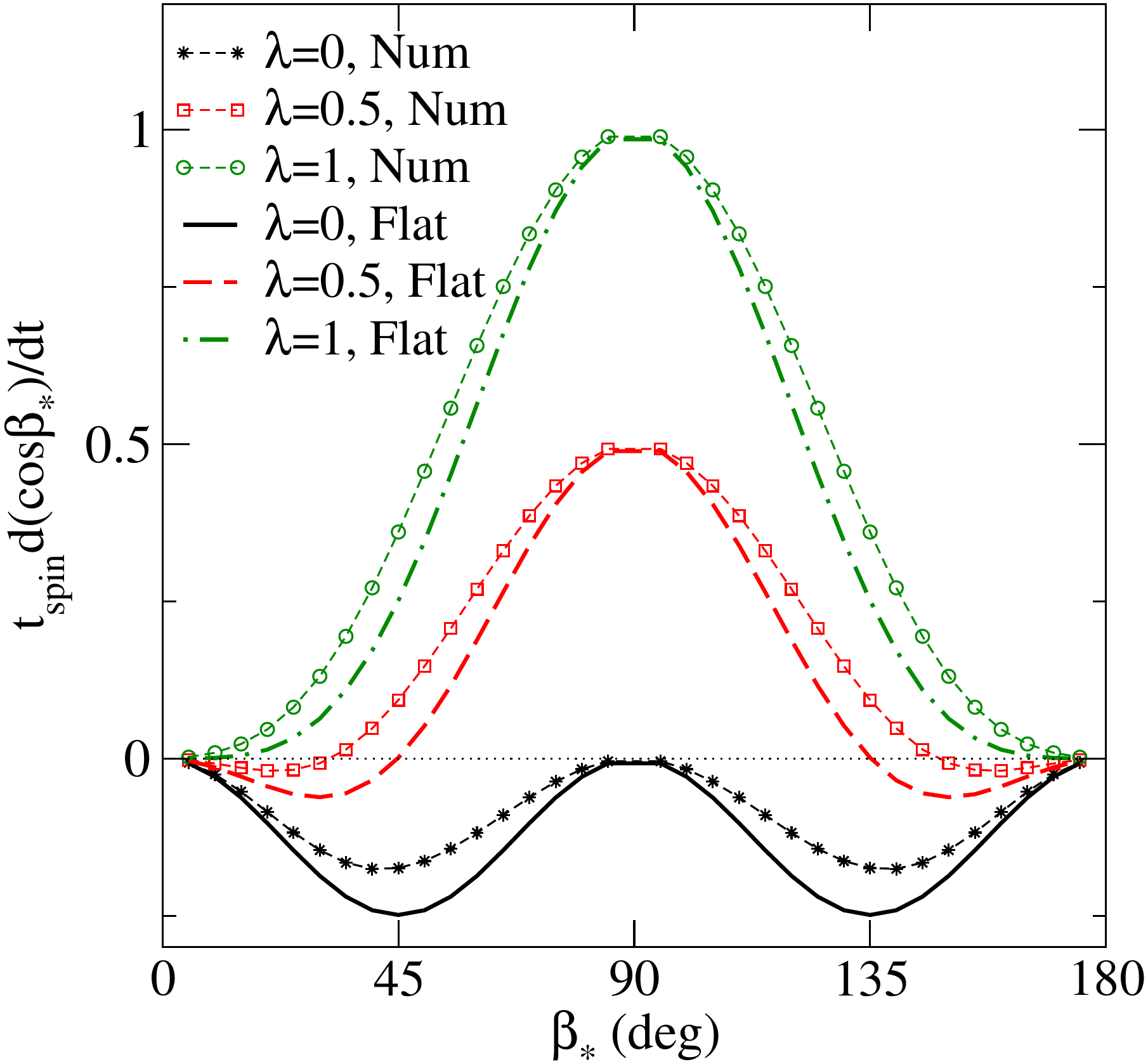}
\caption{Secular evolution rate of the spin-disc inclination angle $\beta_\star$ for discs 
with $\zeta=1$. 
The time derivative of $\cos{\beta_\star}$ is given for the flat-disc approximation (Flat) and for our 
numerical results for warped discs (Num), as well as for 3 different values of the accretion 
parameter $\lambda=0,0.5,1$(see equation~\ref{eq:Nl}). 
The angle $\beta_\star$ will increase if $d\cos{\beta_\star}/dt<0$.}
\label{fig:zeta1seq}
\end{figure}
\begin{figure}
\includegraphics[width=8cm]{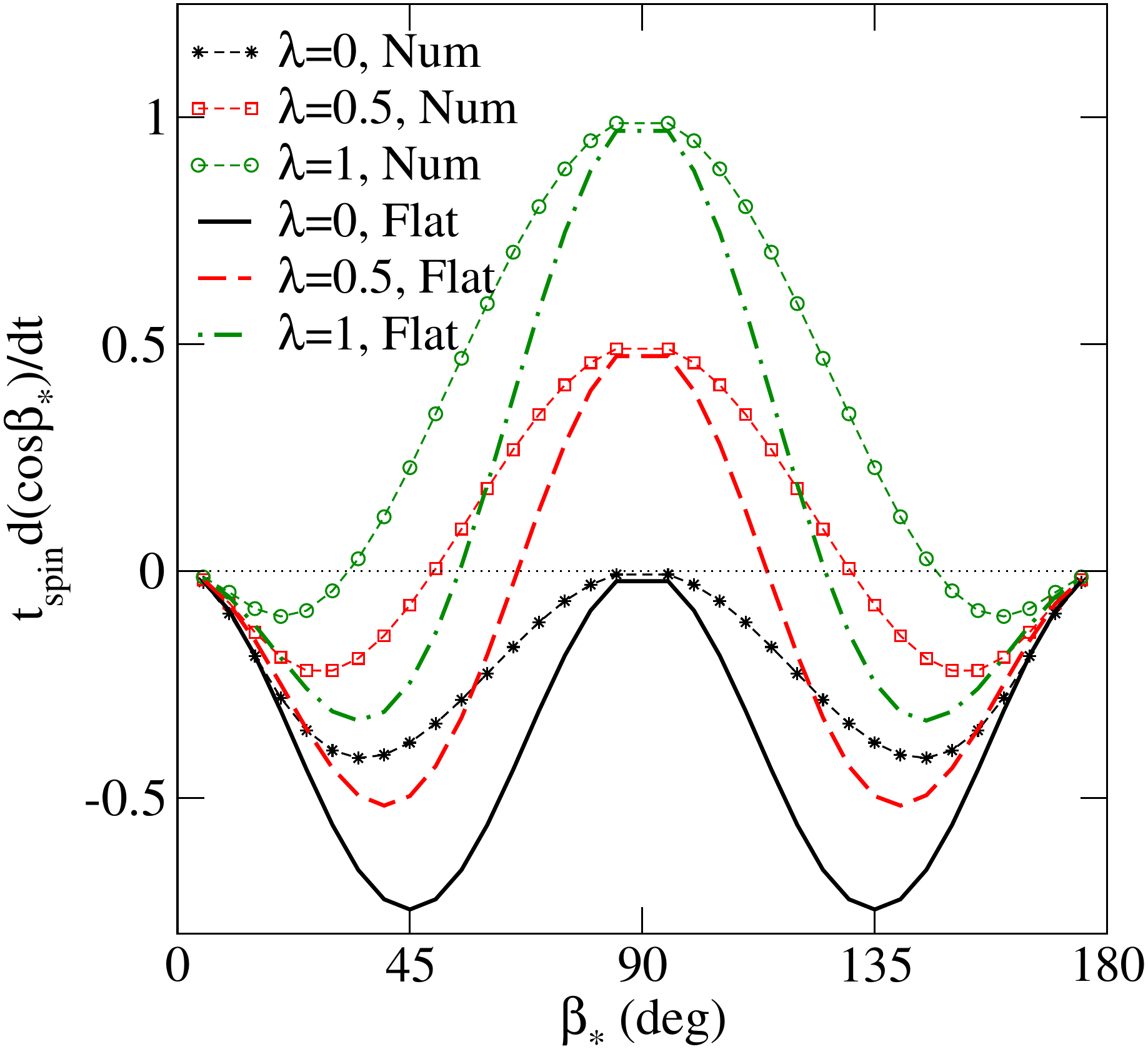}
\caption{Same as Fig. \ref{fig:zeta1seq}, except that we use $\zeta=3$.}
\label{fig:zeta3seq}
\end{figure}
\begin{figure}
\includegraphics[width=8cm]{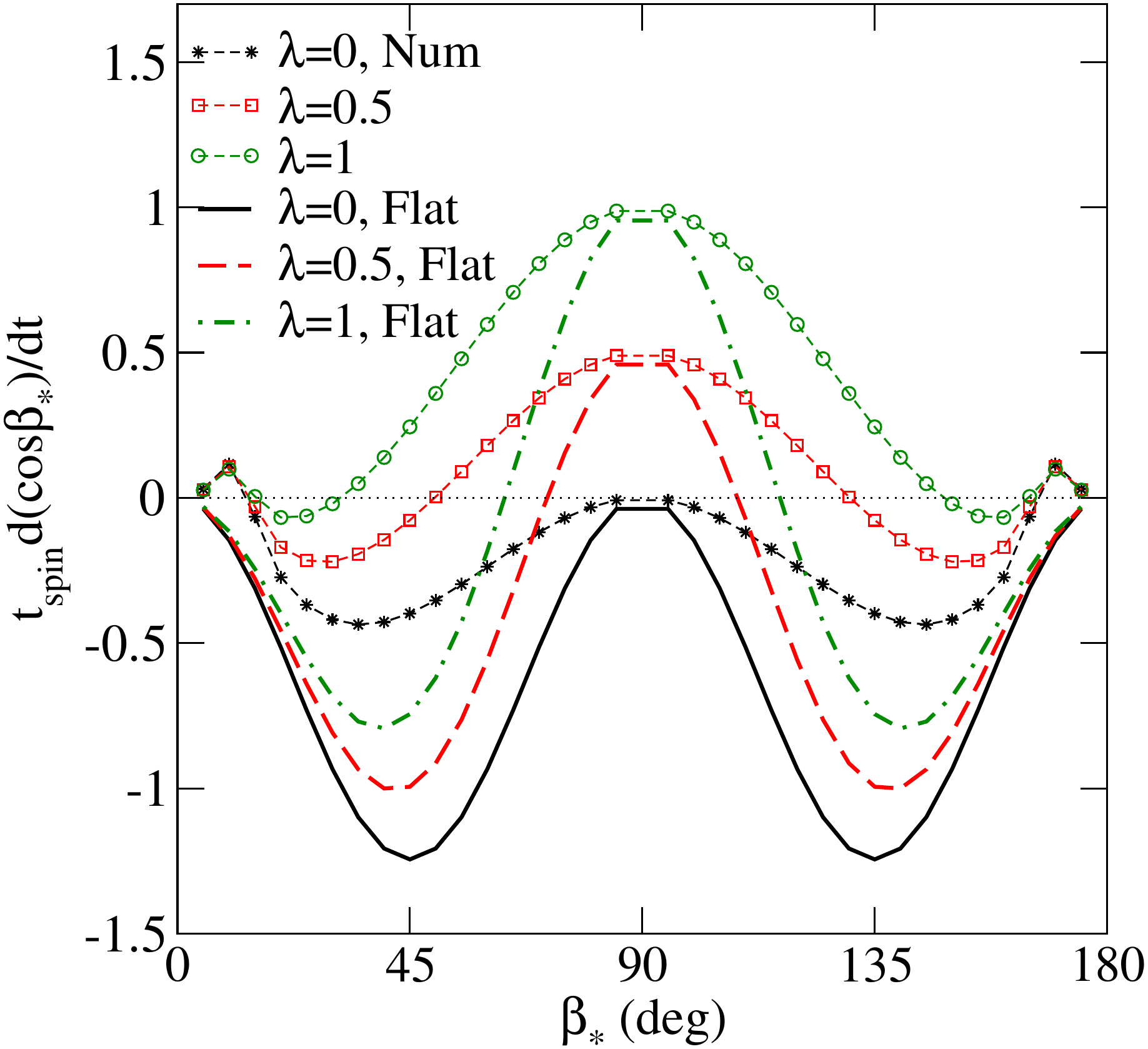}
\caption{Same as Fig. \ref{fig:zeta1seq}, except that we use $\zeta=5$. Note that when the disc is
nearly aligned or nearly anti-aligned, the qualitative behavior of the solution is different from
the predictions of the flat-disc approximation.}
\label{fig:zeta5seq}
\end{figure}

Our current understanding of the effects of magnetic fields close to the inner edge of 
the accretion disc is not sufficient to determine with certainty the range of realistic values of the
parameters $\zeta$ and $\eta$. However, their favored values lie in a region of parameter space where the flat disc approximation appears to hold relatively well ($\zeta \sim 1$, $\eta \sim 0.5$). 
The large deviations from the flat disc model observed at high $\zeta$ are thus unlikely to be
encountered in astrophysical systems, though they cannot be entirely ruled out. Thus, these results
implies that the flat disc approximation is likely to be justified, with the caveat that it tends to
overestimate the value of the misalignment angle $\beta_+$.

\subsection{Varying the precessional torque}
\label{sec:npinf}

The results presented in previous subsections were all obtained with $f=0$, thus 
fixing the choice of the
function $F(\theta_*)$ characterizing the magnetically driven disc precession rate. Using
different values of $f$, even at low $\zeta$ it is possible to find discs which require numerical 
solutions to determine their warp profiles. For example, if we choose $f=1$ instead of $f=0$, the sign and 
magnitude of $\Omega_p$ will change. The twist of the disc becomes more important, so that even 
for $\zeta=1$, $\eta=0.5$, there is a significant deviation from the behavior of the flat-disc 
configuration. Comparisons between the disc profiles for different $f$ can be found in 
Fig.  \ref{fig:ftilt}. The most important feature of these profiles is that, for the larger values of $f$, we have 
a large twist $\gamma(r_{\rm in})$. Hence, the precession term in 
equation~(\ref{eq:warpedspinevol}) 
(proportional to $n_p$), which does not contribute to the evolution of $\beta_\star$ 
in the flat-disc approximation, now has a significant impact. For a twist $\gamma$ such that $\sin
(\gamma-\gamma[r_{\rm out}]) F(\theta_\star)>0$, the precession term directly contributes to the
alignment of the outer disc axis with the stellar spin. This is always the case for disc
twists $|\gamma_{\rm in}| \leq 180^\circ$, as a positive $F(\theta_\star)$ causes the inner disc
to precess in the prograde direction, while $F(\theta_\star) \leq 0$ causes a retrograde precession.
If the precessional torque becomes large enough compared to the warping torque 
(proportional to $n_w$), the long-term evolution of the stellar spin direction will be modified. For our standard parameters and the choices of $\eta=0.5$ and $\lambda=0.5$, we find that discs with 
$f \go 0.5$ will always lead to spin-disc alignment, contradicting the flat 
disc predictions (see Fig. \ref{fig:F5seq}). However, it is worth noting that some configurations with 
high $f$ still allow for
long term misalignments: for example, for $f=0.5$, increasing the strength of the azimuthal B-field 
to $\zeta=3$ leads to a behavior very similar to what we found for $f=0$, $\zeta=3$ (see Fig. \ref
{fig:zeta3seq}), while decreasing the viscosity parameter to $\alpha=0.015$ (and choosing 
$\delta=0.01$) limits the twist of the disc, so that the flat-disc approximation remains valid.

\begin{figure}
\includegraphics[width=8cm]{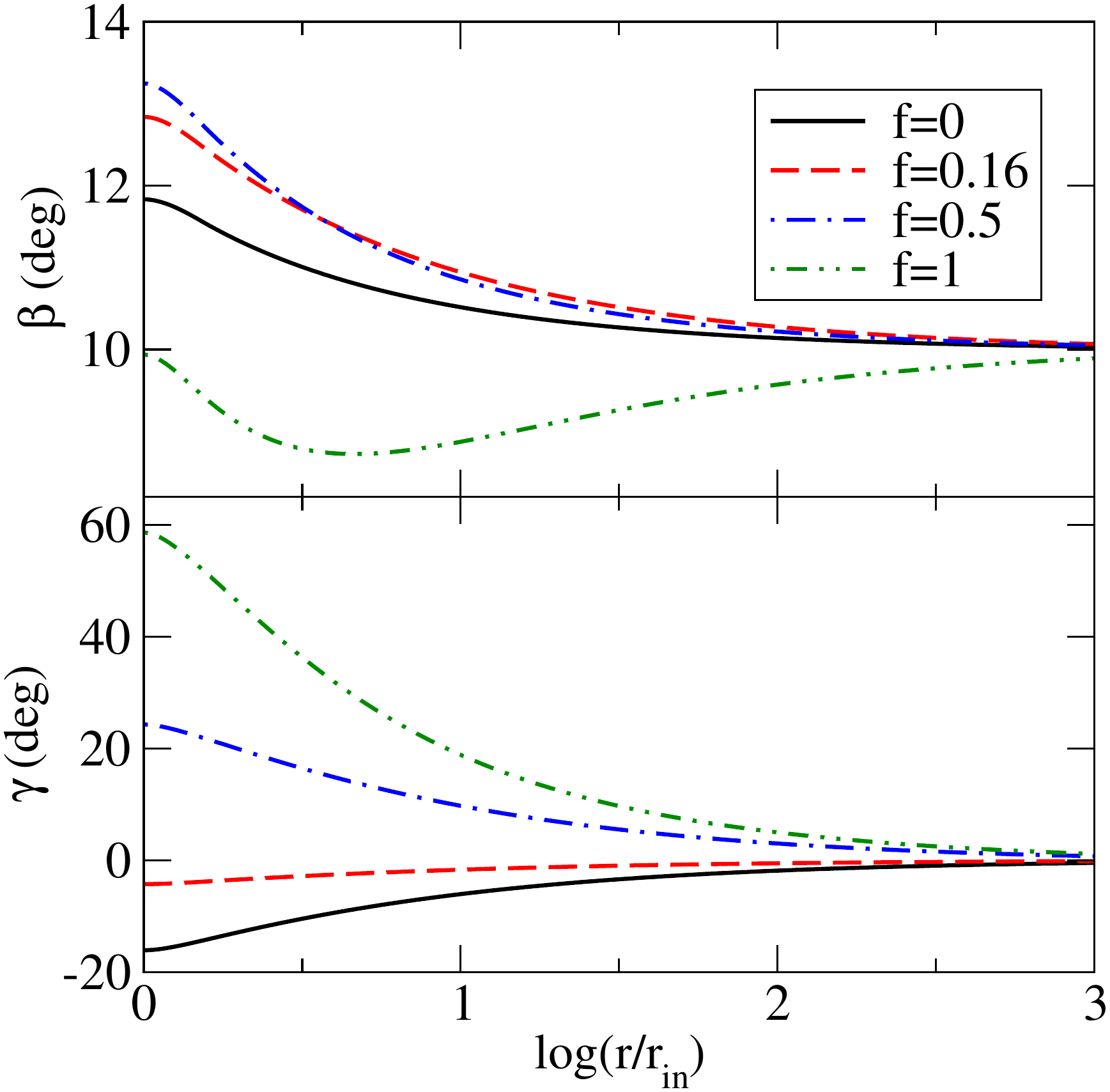}
\caption{{\it Upper panel: }Disc tilt angle $\beta$ for different choices of $f$.
{\it Lower panel: } Twist angle $\gamma$ for the same disc parameters. The outer edge of
the disc is fixed at $r_{\rm out}=10^4r_{\rm in}$.}
\label{fig:ftilt}
\end{figure}
\begin{figure}
\includegraphics[width=8cm]{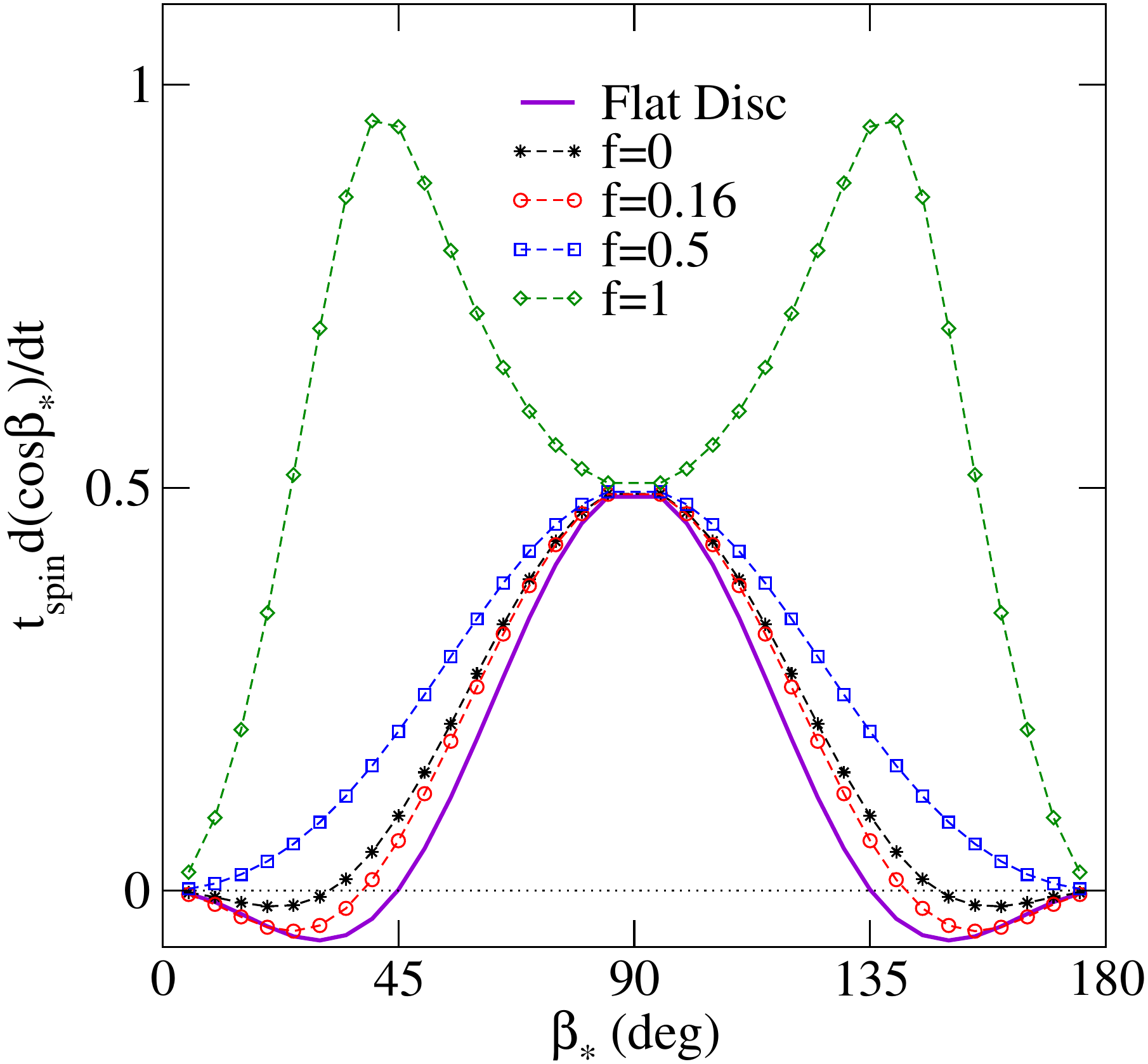}
\caption{Same as Fig. \ref{fig:zeta1seq}, except that we set $\lambda=0.5$ and choose
$f=0,0.16,0.5,1$. For $f=0.16$, the disc twist is very small, and our numerical result matches the
flat-disc approximation better than for $f=0$. For larger values of $f$, the disc twist is large, and
the disc will align with the stellar spin regardless of the initial value of $\beta_\star$.}
\label{fig:F5seq}
\end{figure}

The above results show that the precessional torque can in principle cause non-negligible deviations from the flat-disc model. Nevertheless, for the largest part of the favored parameter
space (small $\alpha$, or large $\alpha$ with small precessional torque), the flat-disc approximation
is justfied.

\subsection{Influence of $Q_3$}
\label{sec:Q3}

As mentioned before, most previous works on warped discs have been done using the formalism
 of \citet{pr1}, which corresponds to $Q_3=0$ in the formalism of \citet{og1}. This is a good approximation, as long as the influence of the small precessional torque due to $Q_3 \neq 0$ is negligible. For the system studied here a small change in the twist of the disc can affect whether a configuration will align over time, or be driven towards a stable misaligned steady-state. In Fig.~\ref{fig:tiltq}, we show the difference in the disc tilt and twist for our standard disc with $\eta=0.5$, using both the formalism of \citet{og1} and \citet{pr1}. Differences in the warp of the disc of a few degrees are observed, though the warps are small in both cases. Because the precessional torque acting on the disc using $Q_3=0$ is smaller, it will be less twisted. This leads to a behavior slightly closer to what the flat-disc approximation predicts. If we choose the accretion parameter  $\lambda=0.5$ (equation~\ref{eq:Nl}), then the flat-disc approximation predicts a  stable misaligned configuration at $\beta_+=45^\circ$. For warp discs, we find that the  misalignment angle is significantly smaller,  $\beta_+=32^\circ$. The
 difference in $\beta_+$ between profiles obtained using $Q_3=3/8$ and $Q_3=0$ is only
 $0.5^\circ$, which is negligible at the level of accuracy our model can achieve. 
\begin{figure}
\includegraphics[width=8cm]{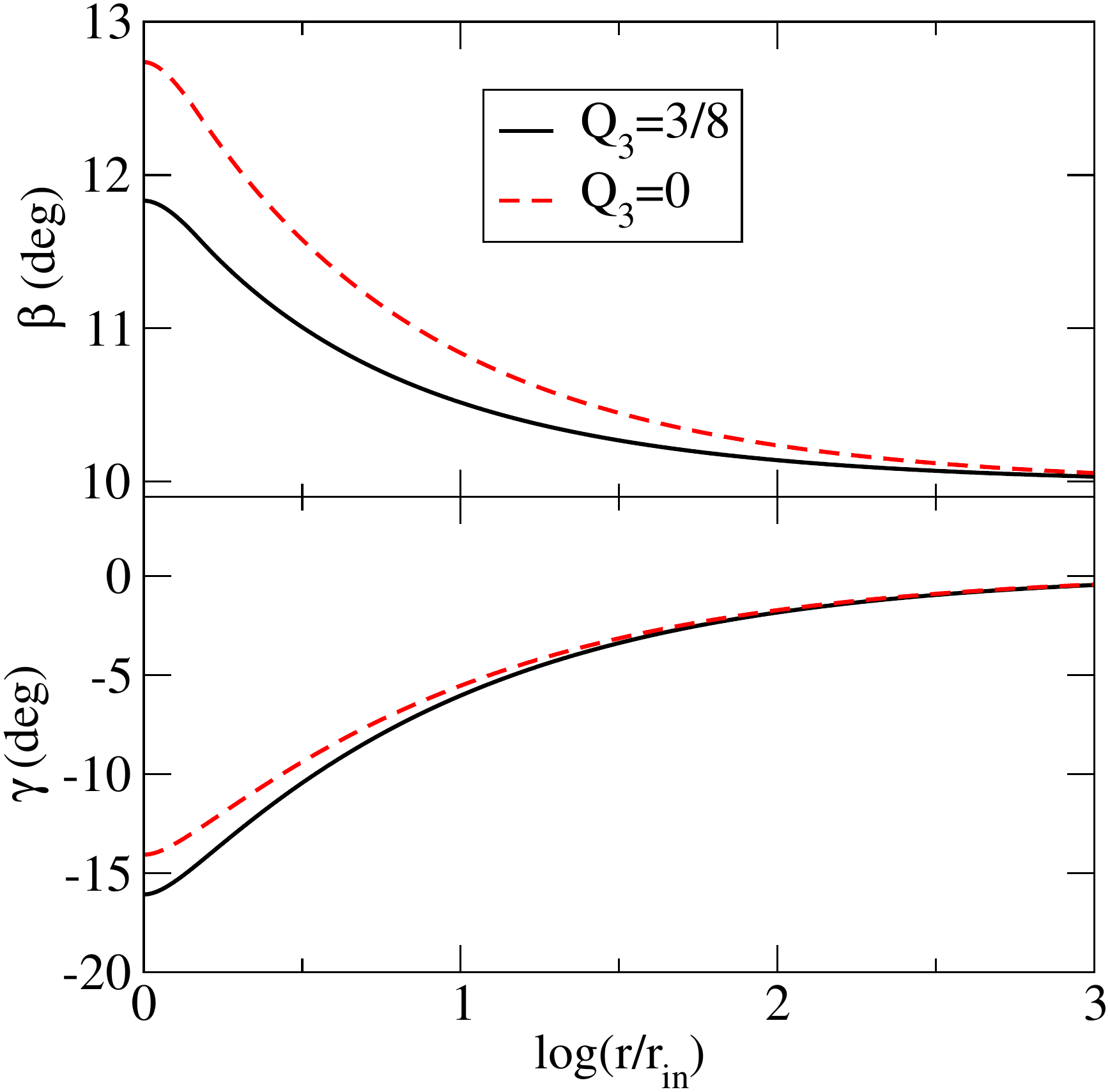}
\caption{{\it Upper panel: }Disc tilt angle $\beta$ for different choices of $Q_3$.
{\it Lower panel: } Twist angle $\gamma$ for the same disc parameters. The outer edge of
the disc is fixed at $r_{\rm out}=10^4r_{\rm in}$.}
\label{fig:tiltq}
\end{figure}

$Q_3$ can also influence the qualitative behavior of the steady-state solutions at high-$\zeta$. For strongly warped discs, it is sometimes possible to have two solutions satisfying the steady-state equations. Choosing $Q_3 \neq 0$ seems to limit the size of the region of parameter space where this happens. For example, for $f=0$, $\eta=0.5$, $\theta_\star=10^{\circ}$ and $\zeta=5.5$, two profiles are acceptable steady-state solutions if we chose $Q_3=0$, while for $Q_3=3/8$ the same parameters lead to a unique solution (see Fig \ref{fig:2sol}).

\begin{figure}
\includegraphics[width=8cm]{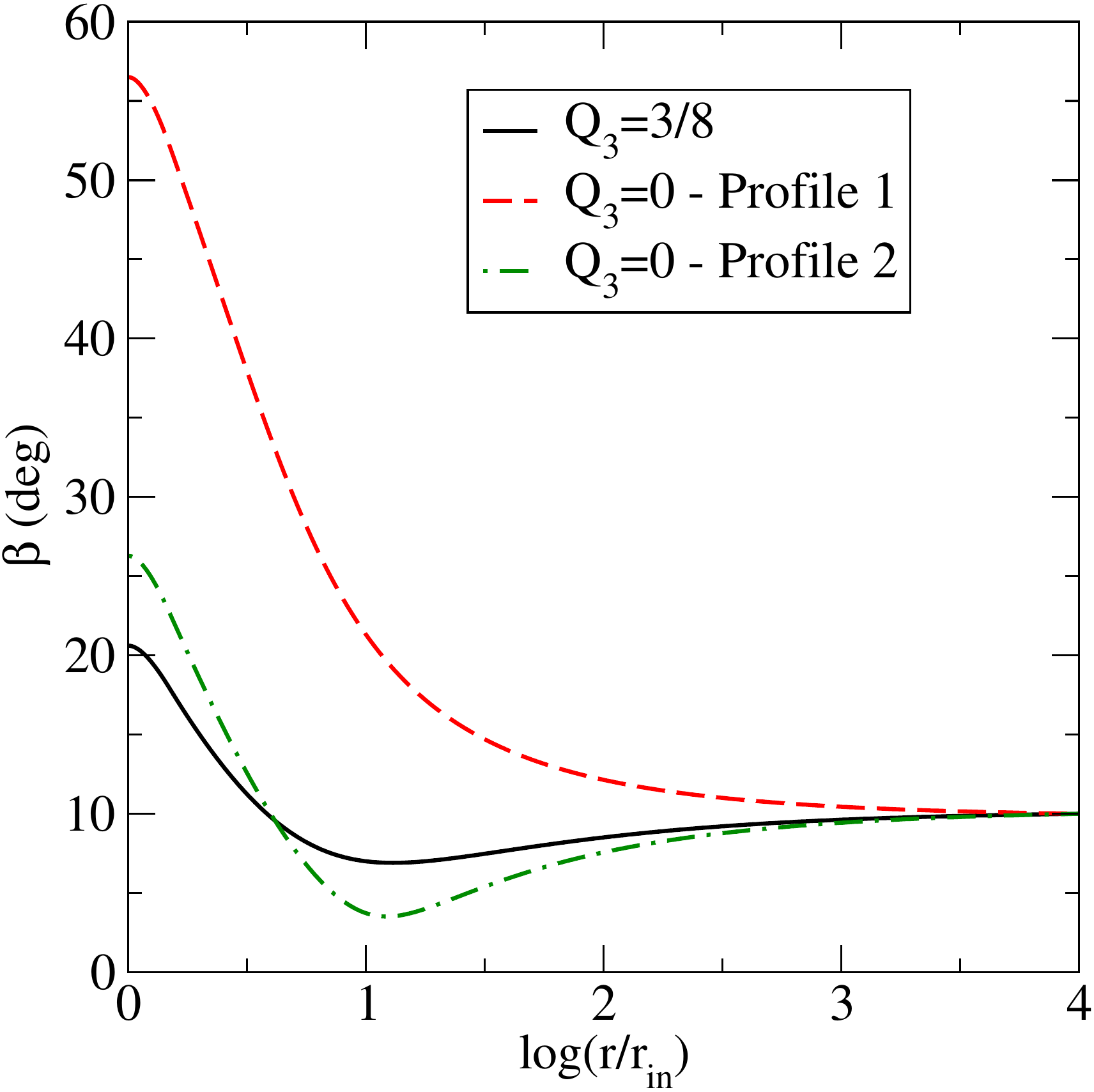}
\caption{Disc tilt angle $\beta$ for $\zeta=5.5$ and $Q_3=0,3/8$. For $Q_3=0$, the steady-state
equations admit two solutions.}
\label{fig:2sol}
\end{figure}

\subsection{Low-Viscosity Discs}

In the linear regime, the equations determining the steady-state profile of the disc are identical
for the $\alpha \leq \delta$ and $\alpha \geq \delta$ cases. In the previous subsections, we have seen that our approximate formulae for the amplitude of the warp, equations~(\ref{eq:deltal}) 
and~(\ref{eq:approxwarp}), give relatively good results for $\alpha \sim \delta = 0.1$. 
We also  confirmed numerically the $\alpha^2$ dependence of the
warp of the disc, shown in Figure~\ref{fig:tiltalpha}.
\begin{figure}
\includegraphics[width=8cm]{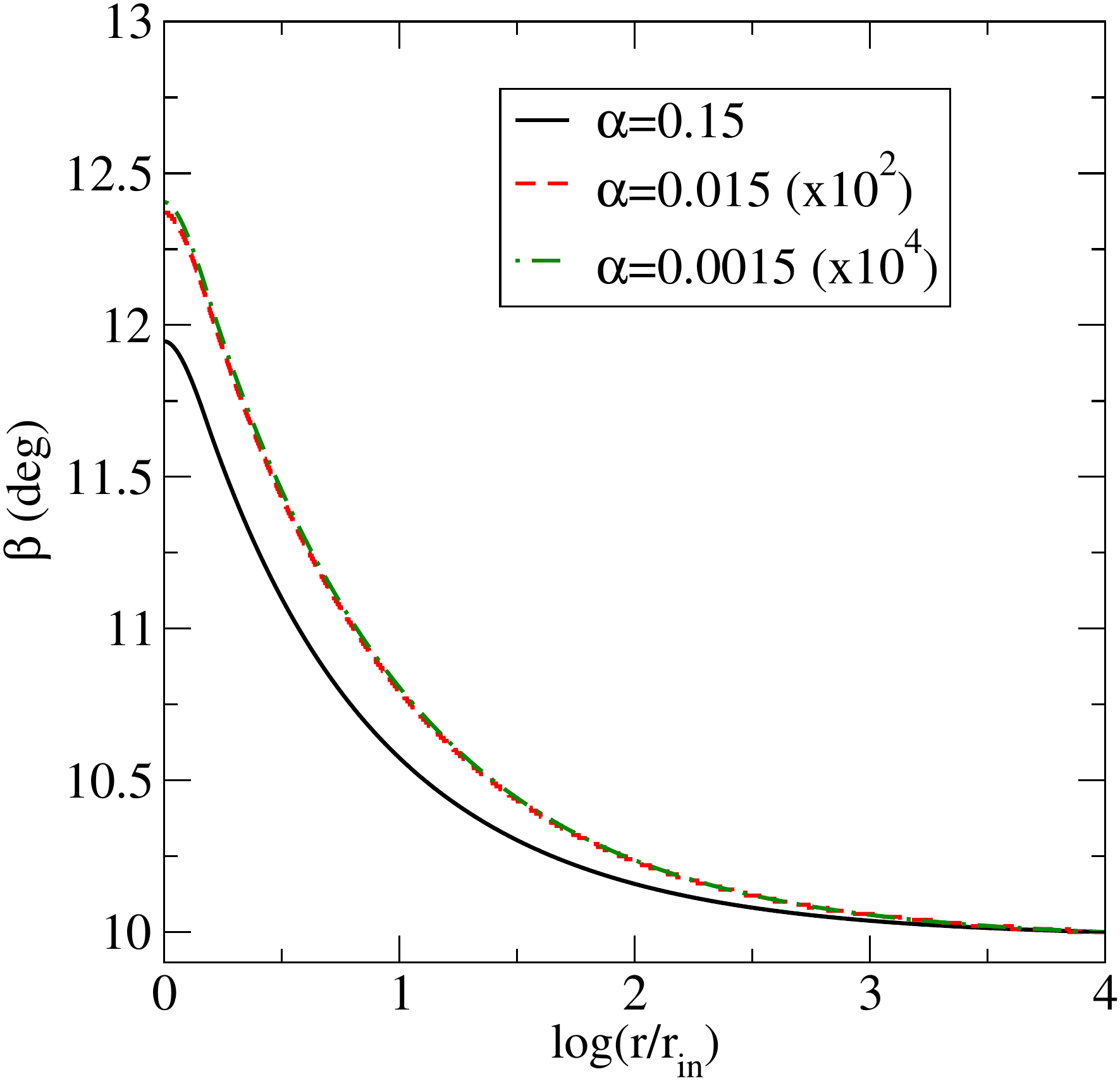}
\caption{Disc tilt angle $\beta$ for $\alpha=0.15,0.015,0.0015$. To check the $\alpha^2$
dependance of $\beta$, the deviation from a flat disc is multiplied by $10^2$ and $10^4$ 
for $\alpha=0.015$ and $\alpha=0.0015$ respectively.}
\label{fig:tiltalpha}
\end{figure}
For smaller viscosities, $\alpha \leq \delta$, we expect the warp to be even smaller, and the linear approximation more accurate. Hence, we can
immediately deduce that the time-averaged warp of low-viscosity discs will be extremely small. For such discs, the flat-disc approximation will nearly always give accurate results for the secular evolution of the stellar spin.

\section{Time Evolution of Disc Warp Toward Steady-State}

Having established the steady-state of warped discs, we now study
their time evolution starting from some generic initial conditions,
when the symmetry axis of the outer disc is misaligned with the
stellar spin. To this end, we evolve
equations~(\ref{eq:dtlambda})-(\ref{eq:dtl}) for high-viscosity discs
and~(\ref{eq:Vwave}) for low-viscosity discs.  Since the timescale to
reach steady state is generally much longer than the local disc
warp/precession time $\Gamma_w^{-1}\sim \Omega_p^{-1}$ [see
  Eq.~(\ref{eq:twarp})], an implicit evolution scheme is
necessary. Our numerical method is detailed in the Appendix.

\subsection{High-Viscosity Discs}
\label{subsec:highvis}

For viscous discs with $\alpha\go \delta=H/r$, we expect the evolution of
the system to occur over the timescale $t_{\rm vis}(r) =
r^2/\nu_2=2\alpha/(\delta^2\Omega)$. 
Note that at the disc inner edge, $t_{\rm vis}(\rin)=(3\alpha^2\zeta
\cos^2\theta_\star/2\eta^{3.5})\Gamma_w^{-1}(\rin)$ [see Eq.~(\ref{eq:tvisgam})] is 
smaller than the warping timescale for typical parameters.
In terms of the dimensionless time $\tau= t/t_{\rm vis}(r_{\rm in})$, 
we expect the disc to reach the steady-state profile
at radius $r$ within a time of order $\tau \sim (r/r_{\rm in})^{3/2}$
(assuming constant $\delta$).
To test this expectation, we 
evolve our standard disc model (see Section 4) for $\eta=0.5$ and
different locations of the outer radius ($r_{\rm out}=100 r_{\rm in}$
and $r_{\rm out}=1000r_{\rm in}$), as well as for a more viscous disc
with $\alpha=0.3$. The disc is initialized in a flat configuration
with $\hatl=\hatl_{\rm out}$ and we
observe its evolution towards the steady-state profile.  In
Figs.~\ref{fig:evolvis1}-\ref{fig:evolvis3}, we plot the disc warp 
profiles at times $\tau=10^{3n/4}$ for $n=0,1,...,4$ --- 
by which point the viscous forces 
should have brought the disc into its steady state up to 
radius $r\sim 10^{n/2} r_{\rm in}$.

\begin{figure}
\includegraphics[width=8cm]{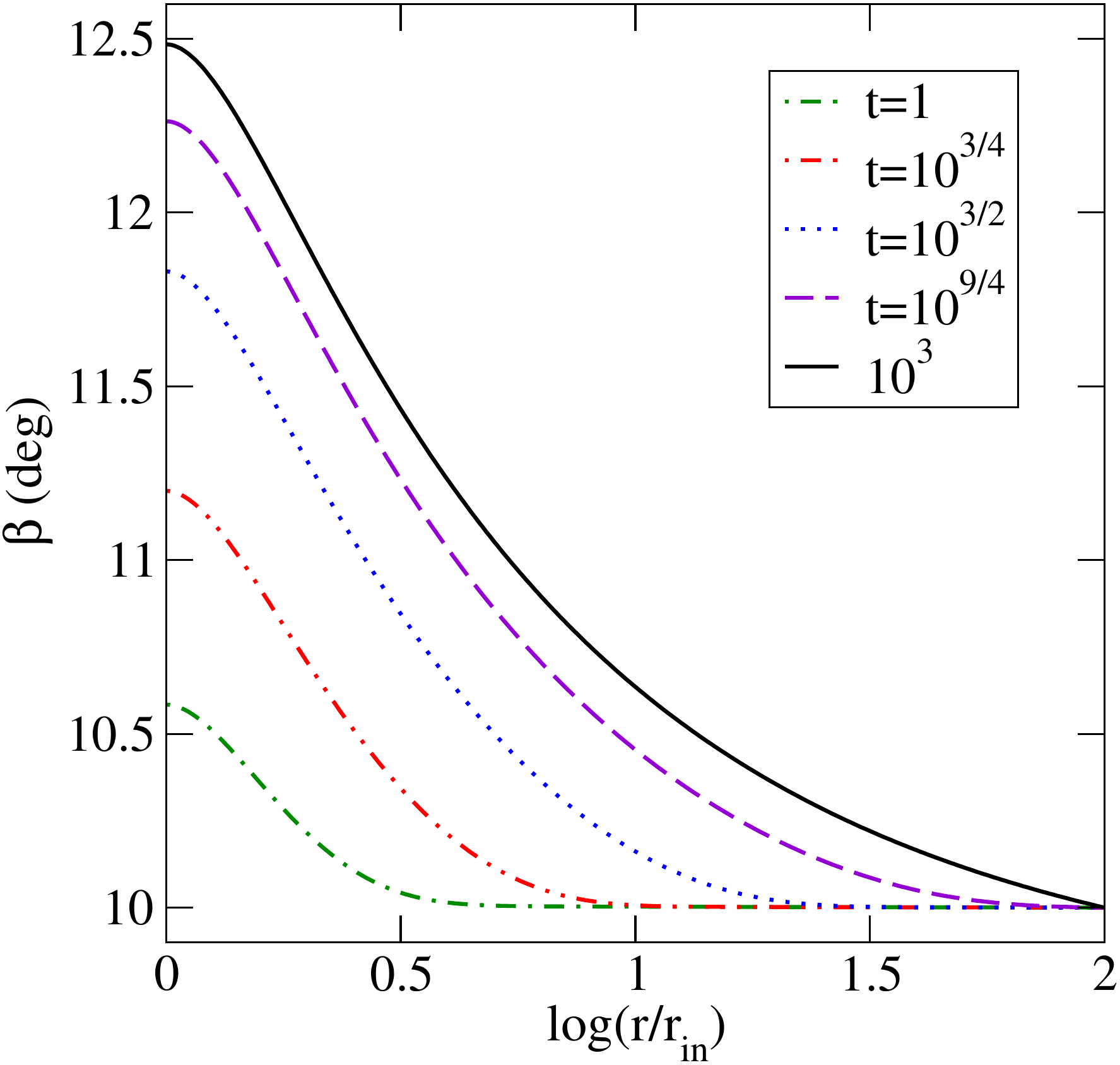}
\caption{Time evolution of the disc tilt angle profile for standard discs with
$\alpha=0.15$ and $r_{\rm out}=100r_{\rm in}$. Time is in units of 
$t_{\rm vis}(\rin)$.}
\label{fig:evolvis1}
\end{figure}

In all cases, we see that the evolution occurs approximately on the
expected timescales: the local distortion of the disc (i.e.
$\partial\hatl/\partial\ln r$) up to radius $r$ does not vary much
past the viscous timescale at that radius. The orientation of the disc
($\hatl$), on the other hand, continues to change to accommodate the
evolution of the disc at larger radii. Overall, the disc will 
reach its equilibrium profile within the viscous timescale 
$t_{\rm vis}(r_{\rm warp})$, where $r_{\rm warp}$ is defined as the
largest radius at which the warp $|\partial\hatl/\partial\ln r|$
is significant.

\begin{figure}
\includegraphics[width=8cm]{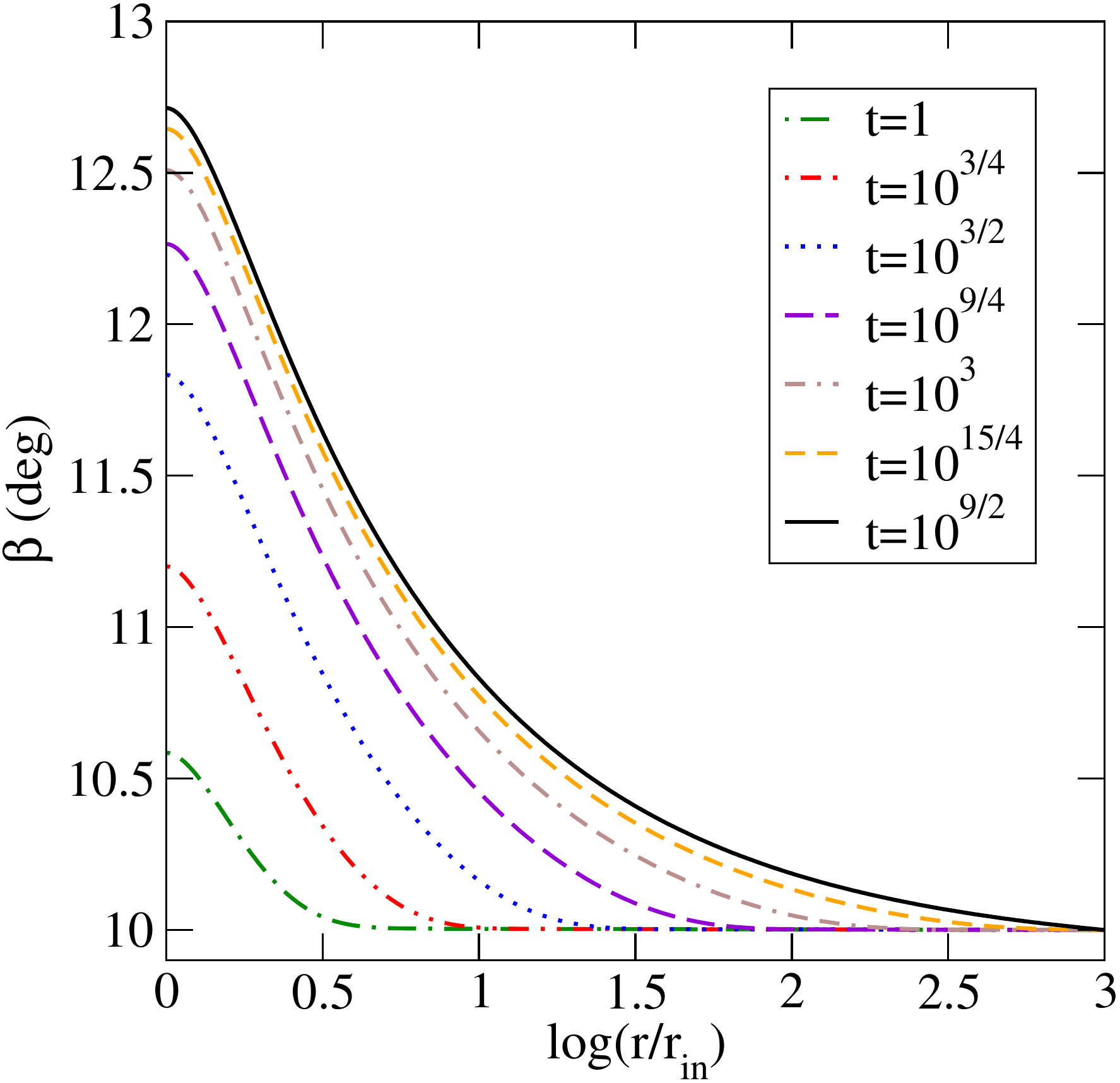}
\caption{Same as Fig.~\ref{fig:evolvis1}, except for $r_{\rm out}=1000r_{\rm in}$.}
\label{fig:evolvis2}
\end{figure}

\begin{figure}
\includegraphics[width=8cm]{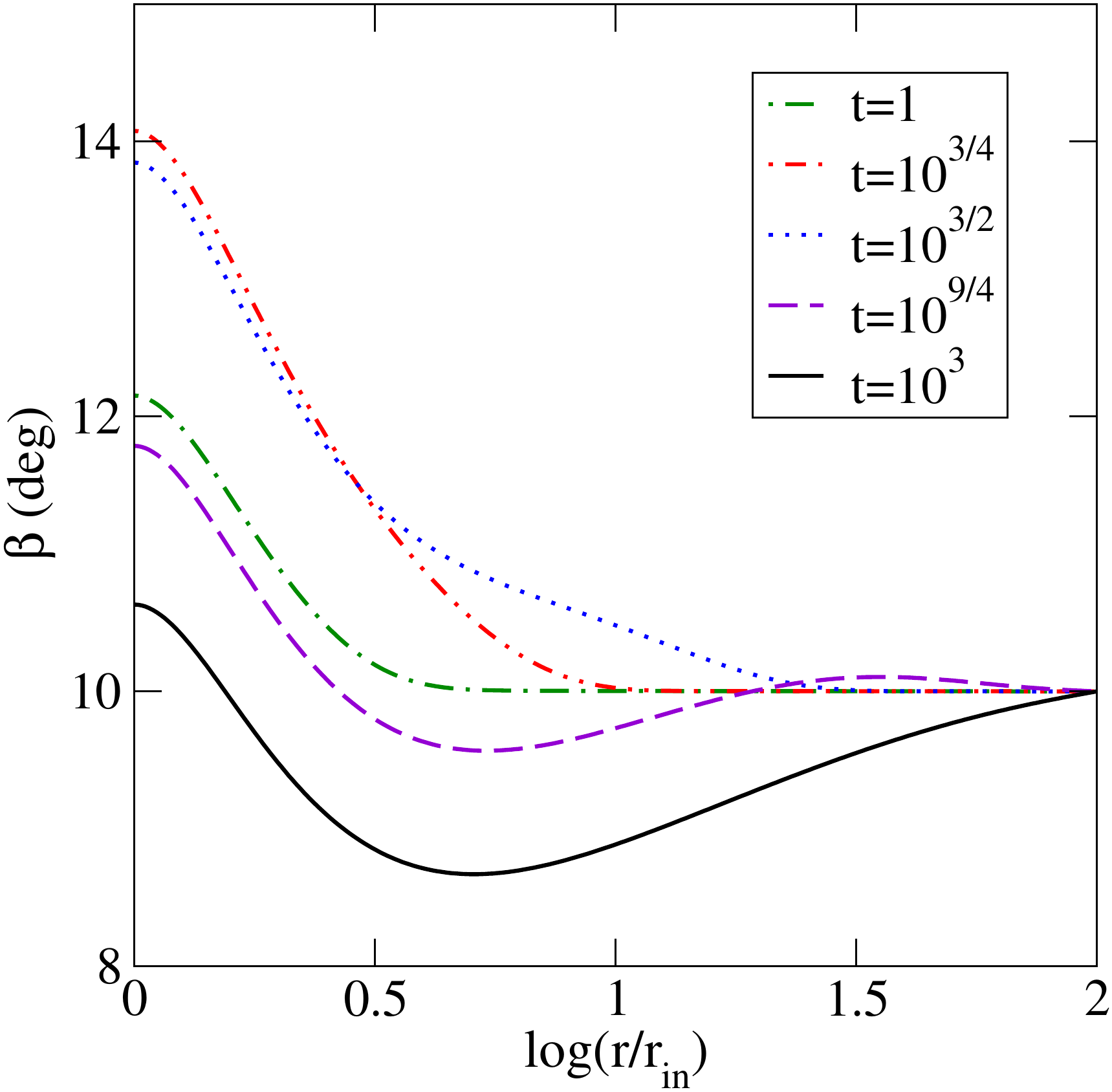}
\caption{Same as Fig.~\ref{fig:evolvis1}, except for $\alpha=0.3$.}
\label{fig:evolvis3}
\end{figure}

For the two simulations with the outer disc boundary at 
$r_{\rm out}=100r_{\rm in}$, we find that $r_{\rm warp} \sim r_{\rm out}$ 
and the disc reaches its steady-state profile within $\tau \sim 1000$. 
At later times, the evolution of the profiles becomes negligible. For the larger disc
($r_{\rm out}=1000r_{\rm in}$), the situation is slightly
different. At $\tau=1000$, the disc has reached its steady-state
distortion up to $r=100r_{\rm in}$. The disc will still evolve up to
$\tau \sim 10^{4.5}$, but as the warp is very small for $r\go 100r_{\rm in}$, 
the changes in the profile are minimal. As most discs studied
in this paper show negligible warps for $r>(10^2-10^3)r_{\rm in}$, we
expect the steady-state to be reached within at most 
\be 
t_{\rm vis}(10^3 r_{\rm in}) \sim 10^{4.5} \left(\frac{r^2}{\nu_2} 
\right)_{\rm in}\sim 500\,\left({\alpha\over 0.15}\right)
\left({\delta\over 0.1}\right)^{-2}~{\rm yrs},
\ee 
regardless of the outer radius of the disc. As this is much
smaller than the evolution timescale for the spin of the star, we are
justified to consider only the steady-state configuration of the disc
when attempting to determine the long-term evolution of the
misalignment between the stellar spin and the orientation of the
outer disc.

\subsection{Low-Viscosity Discs}

The evolution of low-viscosity discs ($\alpha\lo \delta$) is
qualitatively different from high-viscosity discs. 
According to equation~(\ref{eq:Vwave}), perturbations around the steady-state
propagates as bending waves, at roughly half the local
sound speed. Thus, we expect the disc to settle to an equilibrium
within the propagation timescale of these waves,
\ba
\label{eq:twave}
t_{\rm wave} &=& \int_{r_{\rm in}}^{r_{\rm out}} \frac{2dr}{c_s} 
 \sim \frac{4}{3\delta \Omega(r_{\rm out})}\\
 \nonumber
 &\approx&(2\times10^3 {\rm yrs})\frac{0.1}{\delta}
\left(\frac{r_{\rm out}}{100 {\rm AU}}\right)^{3/2}
\ea
In Figures~\ref{fig:wave1}-\ref{fig:wave3}, we show the evolution of
the disc tilt profile $\beta$ as the bending wave propagates across the disc, 
using our standard choice of parameters for the magnetic torques. We consider
different discs: the first two use $\alpha=0.01$ and have their
outer boundaries at $r_{\rm out} = 100 r_{\rm in}$
(Fig.~\ref{fig:wave1}) and $r_{\rm out} = 1000 r_{\rm in}$
(Fig.~\ref{fig:wave2}). The third has a higher viscosity
$\alpha=0.05$, and $r_{\rm out} = 1000 r_{\rm in}$
(Fig.~\ref{fig:wave3}). All three simulations are started from a flat
disc configuration, and show the same behavior: the magnetic torques
perturb the inner disc, and the perturbation propagates outwards over
the timescale $t_{\rm wave}$. Again, this timescale is much less
than the spin evolution timescale.

 \begin{figure}
\includegraphics[width=8cm]{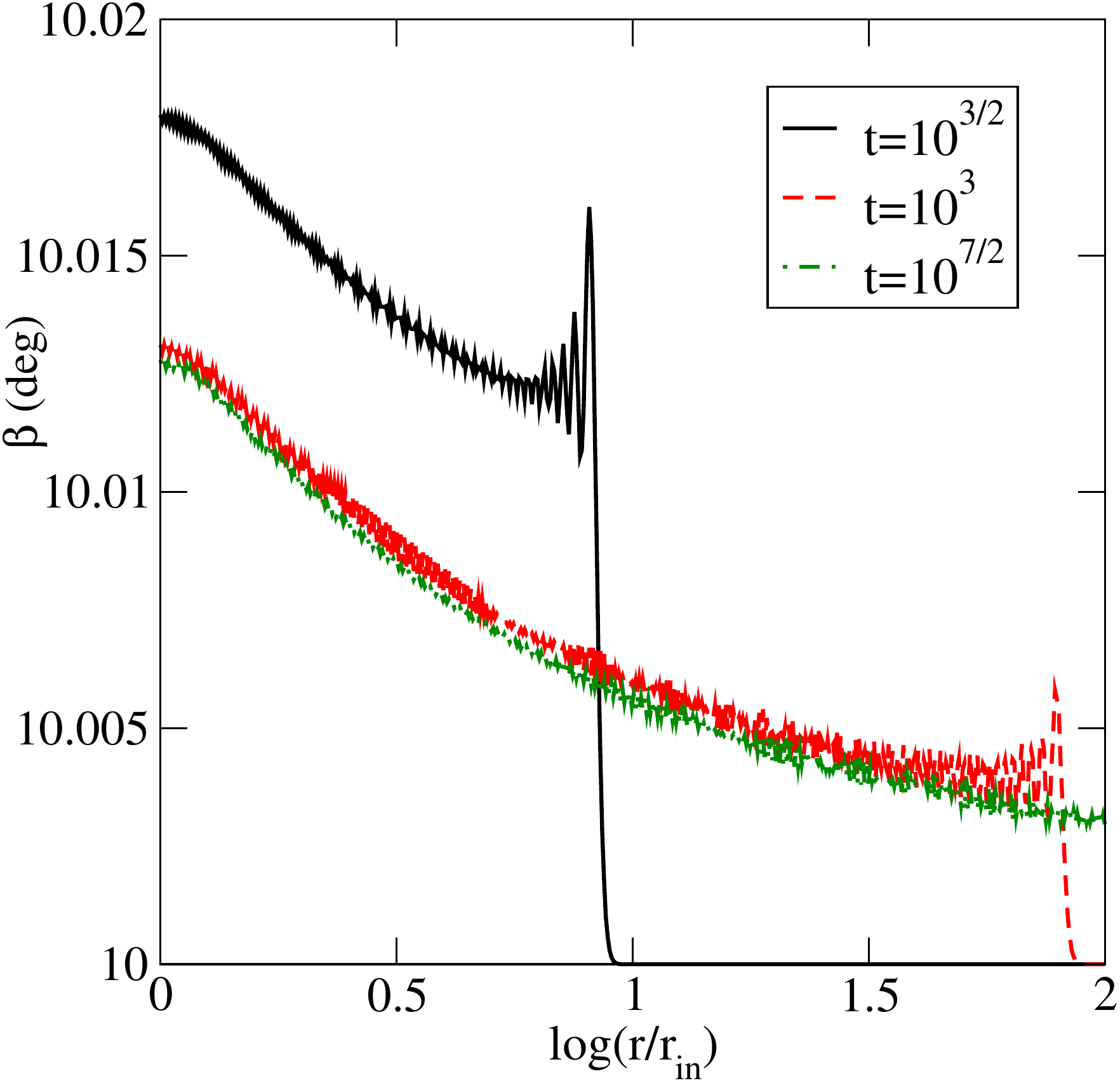}
\caption{Evolution of the disc tilt angle profile for discs with 
$\alpha=0.01$ and $r_{\rm out}=100r_{\rm in}$.
The unit of time is $(\delta \Omega(r_{\rm in}))^{-1}$.}
\label{fig:wave1}
\end{figure}

\begin{figure}
\includegraphics[width=8cm]{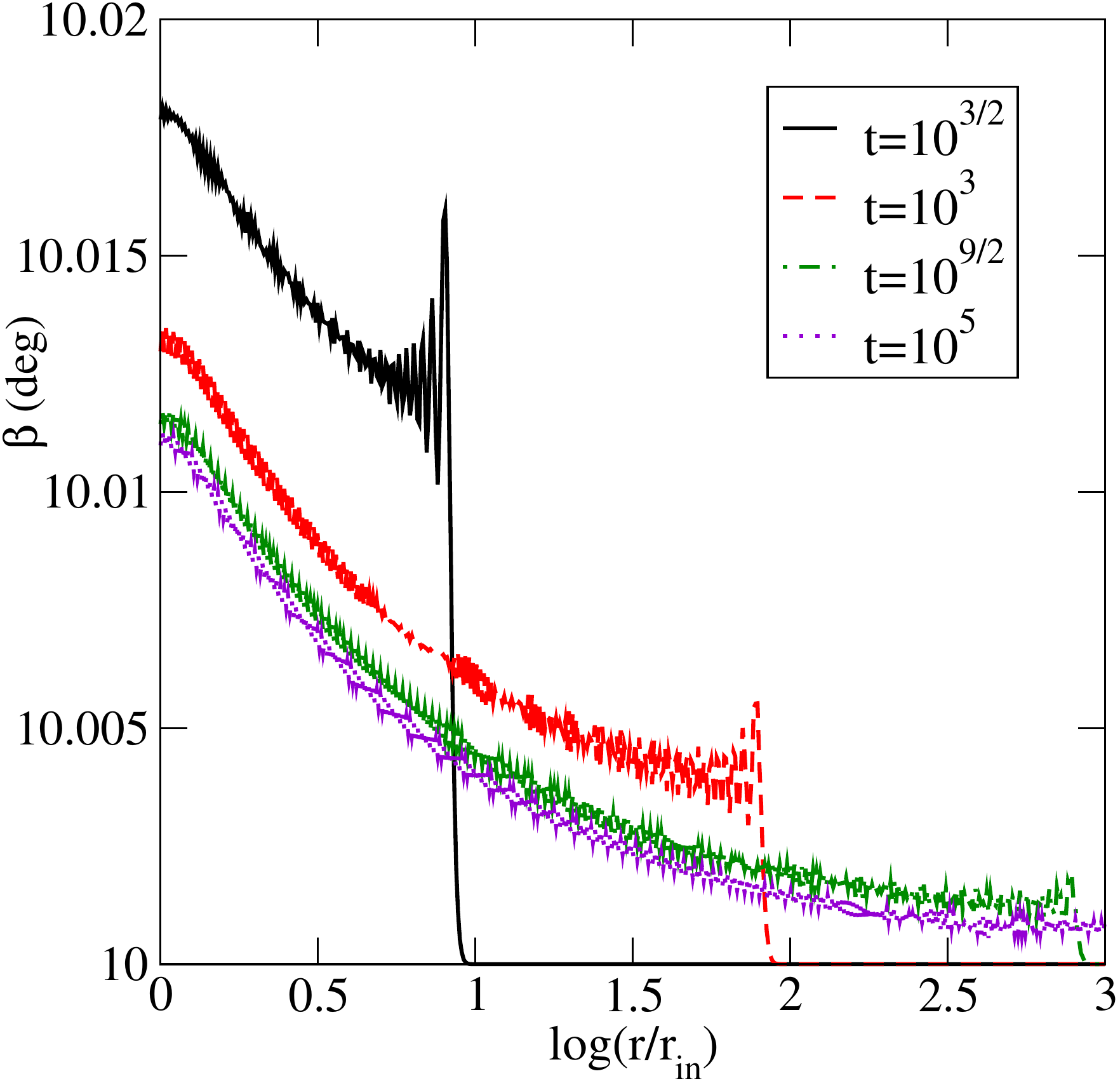}
\caption{Same as Fig.~\ref{fig:wave1} except 
for $r_{\rm out}=1000r_{\rm in}$.}
\label{fig:wave2}
\end{figure}

\begin{figure}
\includegraphics[width=8cm]{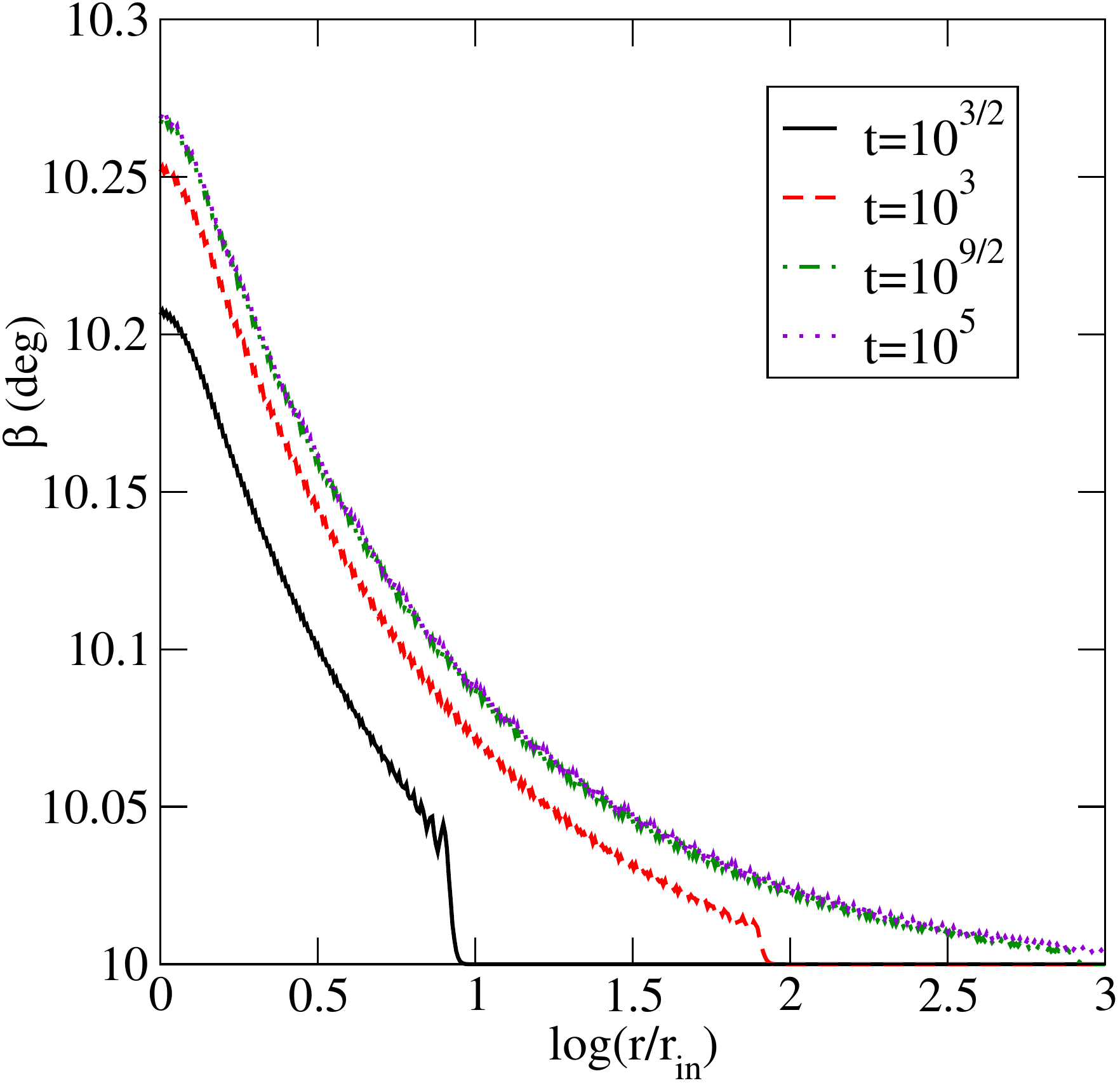}
\caption{Same as Fig.~\ref{fig:wave1} except 
for $\alpha=0.05$ and $r_{\rm out}=1000r_{\rm in}$.}
\label{fig:wave3}
\end{figure}
 
The location of the outer disc radius apparently does not have a
significant influence on the final state of the system. The viscosity,
on the other hand, affects the disc warp amplitude as predicted by
equations~(\ref{eq:deltal}) and~(\ref{eq:approxwarp}): the warp is
proportional to $\alpha^2$, with the amplitude 
$|\beta_{\rm out}-\beta_{\rm in}|$ given 
by Eq.~(\ref{eq:approxwarp}) to within a factor of two.
 

\section{Variations of the Outer Disc orientation}

In the previous section, we have studied the time evolution of warped
discs under the assumption that the orientation of the outer disc is 
fixed. However, a protoplanetary disc is formed inside the star forming core
of a turbulent molecular cloud [e.g., \citet{mo1}].
Thus in general we expect the outer orientation of protoplanetary
discs to have some variations in time. In this section, we study 
how the warped disc and particularly the inner disc orientation respond 
when the outer disc orientation varies by some finite amplitude
(chosen to be $20^\circ$) over a period of time
short compared to the evolution timescale of the disc, and how
such variations affect the secular evolution of the stellar spin direction.

\subsection{High-Viscosity Discs}

\begin{figure}
\includegraphics[width=8cm]{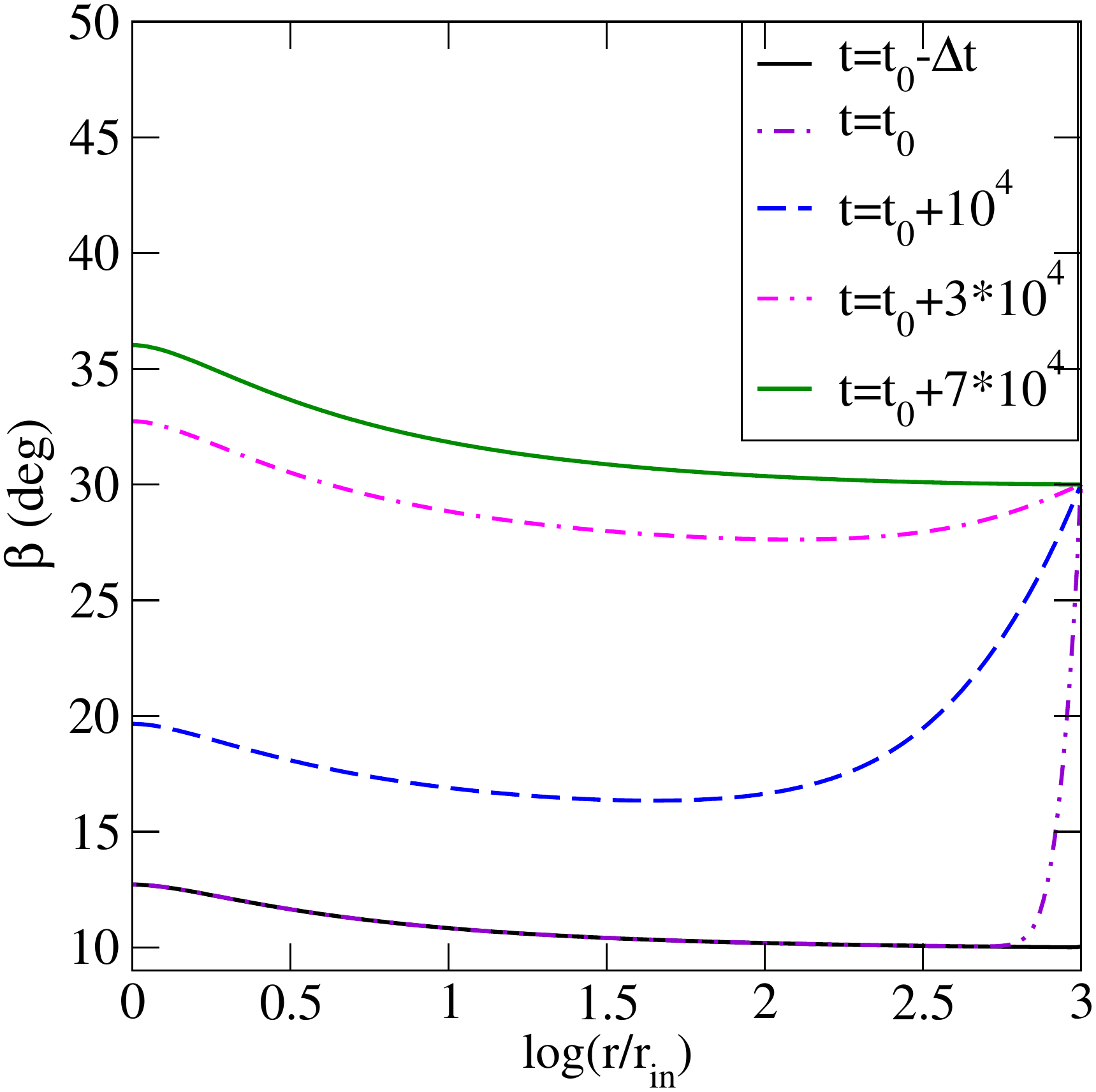}
\caption{Time evolution of the disc tilt angle profile $\beta$ 
for $\alpha=0.15$ and $r_{\rm out}=1000r_{\rm in}$, when  
the outer disc orientation is changed from $\beta(r_{\rm out})
=10^\circ$ at $t=t_0-\Delta t$ to $\beta(r_{\rm out})=30^\circ$ at $t=t_0$,
with $\Delta t=10^3t_{\rm vis}(r_{\rm in})$. 
Time is in units of $t_{\rm vis}(\rin)$.}
\label{fig:visevol}
\end{figure}

We first consider a viscous disc with $\alpha=0.15$ and 
$r_{\rm out}=1000r_{\rm in}$. We choose to vary the outer disc orientation
over $\Delta t=1000 t_{\rm vis}(r_{\rm in}) \sim t_{\rm vis}(100r_{\rm in})$.  
As in the case of the evolution towards the steady-state,
the evolution of the disc occurs on the viscous timescale $t_{\rm vis}$ 
(see Fig.~\ref{fig:visevol}).  However, as significant changes
now take place at the outer radius, the new steady-state configuration
will be reached in a time of order the viscous timescale at the
outer radius $t_{\rm vis}(r_{\rm out})$, whis is larger than $t_{\rm vis}
(r_{\rm warp})$ (see Section \ref{subsec:highvis}).
Nevertheless, even though the steady-state is likely to be reached over
a longer timescale than when the outer orientation is fixed, we still
expect $t_{\rm vis}(r_{\rm out})$ to be significantly less than
the evolution time for the stellar spin $t_{\rm spin}$.
Thus, if the variation of the orientation of the outer disc occurs on a
timescale shorter than $t_{\rm spin}$, the evolution of the stellar
spin is well described by the approximation in which the disc is
assumed to be in its steady-state configuration at all times, and
adapting instantaneously to modifications of its orientation at the
outer boundary.

\subsection{Low-Viscosity Discs}

\begin{figure}
\includegraphics[width=8cm]{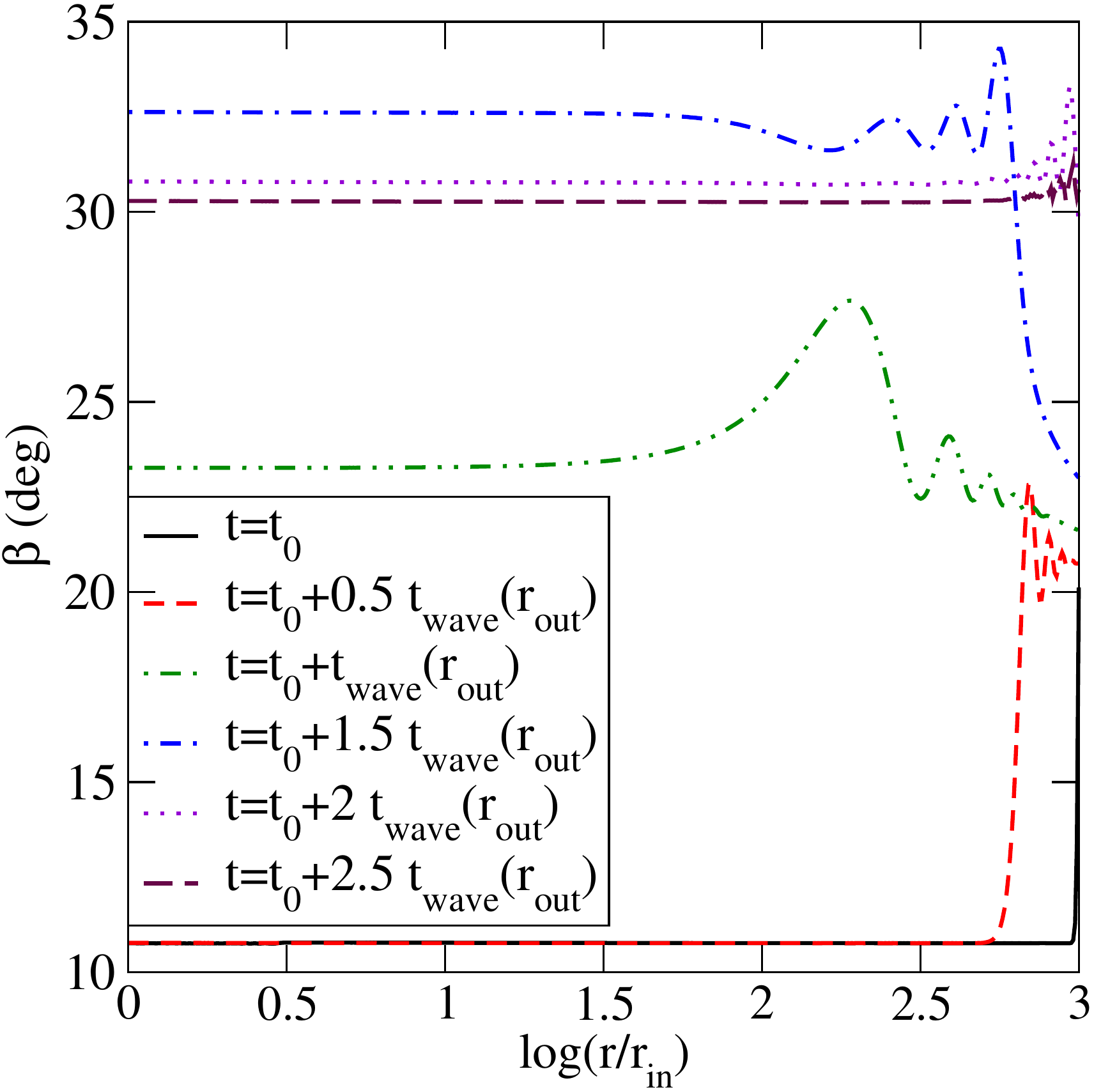}
\caption{Same as Fig.~\ref{fig:visevol} except for 
a disc with $\alpha=0.01$ and $r_{\rm out}=1000r_{\rm in}$,
and the outer disc orientation varies by $20^\circ$ 
over $\Delta t=t_{\rm wave}(100\rin)$.}
\label{fig:lvevol}
\end{figure}

The same type of evolution can also be studied for low-viscosity
discs. If we choose the viscosity parameter $\alpha=0.01$, and change
the orientation of the disc by $20^\circ$ over a timescale $\Delta t =
t_{\rm wave}(100r_{\rm in})$, where $t_{\rm wave}(r)$ is defined by 
equation~(\ref{eq:twave}) with $r_{\rm out}$ replaced by $r$,
we obtain the evolution shown in
Fig.~\ref{fig:lvevol}.  We see that a bending wave created at the
outer boundary propagates inward, until it reaches the inner edge
of the disc where it is reflected. The total time required for the
disc to reach a new steady-state is thus twice the crossing time of
the bending wave, $\sim 8/(3\delta \Omega(r_{\rm out}))$. For
low-viscosity discs, the condition for the steady-state approximation
to be valid when the orientation of the outer disc is allowed to
change over time is thus 
\be 
t_{\rm spin} \go
\frac{8}{3\delta\Omega(r_{\rm out})} \approx 4000 {\rm yrs}.  
\ee 
As $t_{\rm spin} \approx 1 {\rm Myr}$ [see Eq.~(\ref{eq:tspin})], this condition is easily satisfied. 
Also note that the evolution equations of bending waves 
adopted in our analysis are based on the flat-disc approximation. When the outer
disc boundary evolves as fast as shown in Fig.~\ref{fig:lvevol}, this approximation 
is no longer valid. Thus in practice, we should also require $\Delta t \gg 
8/[3\delta\Omega(r_{\rm out})]$.

\section{Application: Anti-aligned exoplanetary orbits}
\label{sec:aa}

Our calculations in Sections 3-5 show that for the most likely 
physical parameters that characterize a magnetic star -- disc system, 
the disc warp is small. Therefore
the long-term evolution of the stellar spin is generally well-described by 
equation~(\ref{eq:flatspinevol}), as long as the orientation of the
outer disc is kept constant. According to~(\ref{eq:flatspinevol}),
three types of spin evolution trend are possible, depending on the parameters of
the system and the initial conditions (Paper I). 
If $\tilde\zeta<\lambda$, the stellar spin and the disc axis will
always align (given enough time) regardless of their initial relative
inclination. If $\tilde\zeta>\lambda$, misalignment between the disc and the
stellar spin will develop, evolving towards one of the two possible 
final states: either  $\beta_\star=\beta_+<90^\circ$, or a perfectly
anti-aligned configuration. The second configuration can only be
reached if the initial disc has a retrograde rotation with respect to the stellar spin, with 
$\beta(t=0) > 180^\circ - \beta_+ = \beta_-$. In this case, to explain the observed
expolanetary systems with retrograde orbits relative to the stellar spin
\citep{triaud}, we have to require that the disc rotates in a very different
direction from the stellar rotation axis during the time of planet formation.
As discussed in paper I [called scenario (2) in Section 5 of Paper I], this
is certainly possible if we consider the complex nature of star formation
in molecular clouds and in star clusters [see also~\citet{blp}].

In Paper I, we describe another potential pathway to create retrograde
exoplanetary systems [called scenario (1)] starting from
prograde-rotating discs. If the disc axis and stellar spin axis are
initially nearly (but not perfectly) aligned, and the magnetic torques are such that the aligned
configuration is unstable, then the misalignment angle will tend towards
$\beta_+$, and no retrograde planets can be produced.
However, this is only true if the orientation of the outer disc does not vary.
If instead we assume that the outer disc experiences a 
change of its orientation $\Delta \beta > \beta_- - \beta_+$ over a timescale 
$\Delta t$ sufficiently long that this change can propagate to the inner disc, 
but short enough that the stellar spin direction does not significantly evolve 
over $\Delta t$, then the star -- disc inclination can jump to 
$\beta>\beta_-$, and continue to evolve towards anti-alignment.
These conditions can be summarized as:
\ba
\Delta \beta &>& \beta_- - \beta_+=180^\circ-2\cos^{-1}\!\sqrt{\lambda\over
\tilde\zeta},\\
t_{\rm disc} & \lo & \Delta t \lo  t_{\rm spin},
\ea
where the disc warp evolution time $t_{\rm disc} \sim t_{\rm vis}(r_{\rm out})$ 
if $\alpha\go \delta$ (high-viscosity disc) and $t_{\rm disc} \sim t_{\rm wave}$ for 
$\alpha\lo\delta$ (low-viscosity disc). As we have seen in Sections 6.1-6.2, 
the second and third conditions are fairly easy to satisfy, as $t_{\rm disc}$ 
is at most of order $10^4~{\rm yrs}$ for a viscous
disc with $r_{\rm out} \sim 10^4r_{\rm in}$ (and $t_{\rm disc}$ 
would be significantly shorter for a smaller outer disc radius), 
while $t_{\rm spin} \sim 10^6 {\rm yrs}$ for typical parameters
[See Eq.~(\ref{eq:tspin})]. The potential to satisfy
the first condition, on the other hand, will 
depend on the fraction of the disc angular momentum which is
accreted by the star (the parameter $\lambda$ in 
equation~\ref{eq:Nl}) and the magnetic warp efficiency (the parameter
$\tilde\zeta$).
If the star only accretes a small fraction of the angular momentum 
($\lambda\ll 1$), then the angles $\beta_{\pm}$ are both close to $90^\circ$, 
and small variations of the outer disc are sufficient
to allow the system to jump to the retrograde state and eventually evolve towards
the anti-aligned configuration. 

\section{Discussion}

The main finding of our paper is that although magnetic interactions
between a protostar and its disc have a strong tendency to induce 
warping in the inner disc region, internal stresses in the disc
tend to suppress the warping under most circumstances. The result is that
in steady-state, the whole protoplanetary disc approximately lies
in a single plane, which is determined by the disc angular momentum at large radii
(averaging out the dynamical warps which vary on timescales of order the stellar rotation period ---
such dynamical warps do not affect the secular evolution of the stellar spin).
The reason for the small steady-state disc warp is that 
the effective viscosity acting to suppress disc warp, 
$\nu_2\simeq \nu_1/(2\alpha^2)$, is much larger than the viscosity 
($\nu_1=\alpha H c_s$) responsible for angular momentum transfer within the
disc \citep{pp1,og1}. In fact,
our anaylsis of the steady-state magnetically driven disc warp shows that,
in the linear regime, the disc inclination angle (relative to the 
stellar spin axis) varies from the outer disc to inner disc by the amount 
[see Eqs.~(\ref{eq:cosbeta}) and (\ref{eq:approxwarp})]
\be
|\beta_{\rm in}-\beta_{\rm out}|\sim 
\left(t_{\rm vis}\Gamma_w\sin 2\beta\right)_{\rm in}
\sim {\alpha^2\zeta\sin (2\beta_{\rm in})\over\eta^{7/2}}.
\ee
where $t_{\rm vis}=r^2/\nu_2$ is the viscous time and
$\Gamma_w$ is the warping rate due to the magnetic torque.
This result is valid regardless of whether the warp perturbations
propagate diffusively (for $\alpha\go H/r$, high-viscosity discs) or as bending waves
(for $\alpha\lo H/r$, low-viscosity discs). Thus, for the preferred values of the parameters 
$\eta\sim 0.5$, $\zeta\sim 1$, we find 
$|\beta_{\rm in}-\beta_{\rm out}|\ll 1$ for $\alpha\ll 0.3$.
Moreover, our analysis of the time evolution of warped discs shows
that, starting from a generic initial condition, the steady-state 
can be reached quickly, on a timescale shorter than the characteristic timescale
for the evolution of the stellar spin orientation.

Overall, our study of magnetically driven warped discs presented in
this paper justifies the approximate analysis (based on the flat-disc
approximation) of the long-term evolution
of spin-disc misalignment presented in Paper I. Nevertheless,
we note that even relatively small disc warps can modify the 
``equilibrium'' spin -- disc inclination angles $\beta_\pm$
(see Fig.~1) from the flat-disc values, thereby affecting the 
``attractors'' of the long-term evolution of the 
spin -- disc inclination angle. If we allow for more extreme
parameters (but still reasonable by physical considerations) 
for the disc -- star systems, much larger disc warps become
possible and qualitatively different evolutionary trends
for $\beta$ may be produced (see Figs.~\ref{fig:zeta1seq}-\ref{fig:zeta5seq}
and~\ref{fig:F5seq}).

Taken together, the results of this paper and paper I demonstrate that
at the end of the first stage of the planetary system formation (see
Section 1), the inclination angle between the stellar spin and the
angular momentum axis of the planetary orbit may have a wide range of values,
including alignment and anti-alignment (see also section~\ref{sec:aa}).  Dynamical processes (e.g.,
planet-planet scatterings and Kozai interactions) in the second stage,
if they exist, would further change the spin -- orbit misalignment angle.
More work is needed to determine the relative importance of the two
stages in shaping the properties of planetary systems.  Currently, the
orbital eccentricity distribution of exoplanetary systems
suggests that the second stage is important (e.g., Juric \& Tremaine
2008). On the other hand, as noted in paper I, the $7^\circ$ misalignment between the ecliptic plane of the solar system and the sun's equatorial plane may be explained by the magnetically
driven misalignment effect studied in this paper. Also, the
recent discovery of Kepler-9 (Holman et al.~2010), a planetary system
with two or three planets that lie in the same orbital plane, seems to
suggest that at least some planetary systems are formed in a ``quiet''
manner without violent multi-body interactions.  Obviously, measuring
the stellar obliquity of such ``quiet'' systems would be most
valuable.

\section*{Acknowledgments}

DL thanks Doug Lin, Gordon Ogilvie and other participants of the KITP 
Exoplanet program (Spring 2010) for useful discussions, and
acknowledges the hospitality of the Kavli Institute for Theoretical
Physics at UCSB (funded by the NSF through Grant PHY05-51164).
FF thanks Harald Pfeiffer for useful discussions on the numerical evolution of warped
discs, as well as for access to his evolution code for comparison tests.
We thank the referees for useful comments which improved the paper.
This work has been supported in part by NASA Grant No NNX07AG81G and NSF
Grant No AST 1008245.

\appendix

\section{Numerical Method for Solving Warp Evolutions}

We evolve equations~(\ref{eq:dtlambda}-\ref{eq:dtl}) for viscous discs and~(\ref{eq:Vwave})
for low-viscosity discs with an implicit Crank-Nicholson evolution algorithm
inspired by the method used by~\citet{pl1} to study the behavior of a disc accreting
onto a magnetic star when the orientation of the outer disc $\hatl(r_{\rm out})$ is aligned with
the stellar spin $\hat\bomega_s$. The evolution equations are all of the form
\be
\label{eq:genev}
\frac{\partial}{\partial \tau} y = A(x,y) \frac{\partial^2}{\partial x^2}y + B(x,y) \frac{\partial}{\partial x}y 
+ C(x,y),
\ee
and are discretized at the N vertices $x_{0,1,...,N-1}$ of our numerical grid as
\ba
\frac{\partial}{\partial \tau} y(x_i) &=& \frac{\tilde y_i - y_i}{\Delta \tau}\\
A(x,y) \frac{\partial^2}{\partial x^2}y(x_i) &=& \tilde A_i \frac{\tilde y_{i+1}+\tilde y_{i-1}-2 \tilde y_i}{2\Delta x^2}  +\\
\nonumber
&& A_i \frac{y_{i+1}+ y_{i-1}-2 y_i}{2\Delta x^2}\\
B(x,y) \frac{\partial}{\partial x}y(x_i) &=& \tilde B_i \frac{\tilde y_{i+1}-\tilde y_{i-1}}{4\Delta x}
+B_i \frac{y_{i+1} - y_{i-1}}{4\Delta x}\\
C(x,y)&=& \frac{1}{2} (\tilde C_i + C_i)
\ea
where $\tilde y_i$ is the value of $y$ at point $x_i$ and time $\tau + \Delta \tau$. At each time step
of the Crank-Nicholson algorithm, we start from an initial guess for $\tilde y_i$ obtained by 
extrapolating from the three previous time steps. From that guess $\tilde y^0_i$, we evaluate 
$\tilde A$, $\tilde B$ and $\tilde C$. Assuming these functions as fixed, we can then obtain
$\tilde y_i$ by solving a tridiagonal system of equations. This gives us an updated guess 
$\tilde y^1_i$ for the value of the function at $\tau + \Delta \tau$. We then repeat the operation
until the step $s$ for which the condition
\be
\max_i{|y^{s}_i-y^{s-1}_i|}< \epsilon_{\rm tri}
\ee
is satisfied for some chosen tolerance $\epsilon_{\rm tri}$.

The main advantage of this implicit method is that the time step $\Delta \tau$ can be much larger
than the Courant limit when the variable $y$ evolves slowly in time. In practice, $\Delta \tau$ is
chosen so that the condition
\be
\max_i{|y^{s}_i-y^{0}_i|}< \epsilon_{\rm CN}
\ee
is satisfied for $\epsilon_{\rm tri} \ll \epsilon_{\rm CN} \ll 1$. We choose 
$\epsilon_{\rm CN} \sim 10^{-4}$ in our simulations (the Crank-Nicholson algorithm is 
second-order convergent in time, and we verified both the convergence and the fact that
we could obtain sufficient precision for that choice of $\epsilon_{\rm CN}$). In order to limit
the computational cost of each time step, we also modify $\Delta \tau$ so that we only need 
$s_{\rm obj}$ tridiagonal solves for each Crank-Nicholson time step (in our simulations,
$s_{\rm obj}=14$).

To evolve the disc-magnetic star system, we also need to choose an implementation of the
inner and outer boundary conditions. We encounter two types of boundary conditions:
Dirichlet conditions of the type $y=y_{\rm BC}$ are enforced by replacing the discretized version
of~(\ref{eq:genev}) by $\tilde y_{0,N-1} = y_{\rm BC}$, while Neumann conditions of the type
$y'=y'_{\rm BC}$ are enforced by explicitly replacing $y'$ by $y'_{\rm BC}$ whenever necessary
in~(\ref{eq:genev}). If a second derivative is required to evaluate~(\ref{eq:genev}) at the boundary,
$y_{-1}$ and $y_N$ are obtained using $(y_{i+1}-y_{i-1})=(\Delta x) y'_i$ and the known value of
$y'$ at the boundary.



\end{document}